\documentclass[11pt, a4paper]{article}
\pdfoutput=1

\usepackage{amsmath}
\usepackage{amsfonts}
\usepackage{amssymb}
\usepackage{graphicx}
\usepackage{mathrsfs}
\usepackage{latexsym}
\usepackage{color}
\usepackage{cite}
\usepackage{slashed, cancel}
\usepackage{hyperref}

\numberwithin{equation}{section}

\definecolor{rossos}{rgb}{0.8,0.2,0.3}
\definecolor{bluscuro}{rgb}{0.15, 0.2, .85}
\definecolor{bluchiaro}{cmyk}{1,.3,0.,0.1}
\hypersetup{colorlinks, citecolor=bluscuro, linkcolor=bluscuro, urlcolor=bluscuro}

\makeatother   % Cancel the effect of \makeatletter
\newcommand{\GeV}{{\rm \,GeV}}
\newcommand{\TeV}{{\rm TeV}}

\def\de{\textrm{d}}

 \def\be   {\begin{equation}}   \def\ee   {\end{equation}}
 \def\ba   {\begin{array}}      \def\ea   {\end{array}}
 \def\bea  {\begin{eqnarray}}   \def\eea  {\end{eqnarray}}
 \def\bean {\begin{eqnarray*}}  \def\eean {\end{eqnarray*}}
 \def\nn{\nonumber}

\begin{document}

\begin{flushright} 
SISSA  04/2014/FISI
\end{flushright}

\vspace{0.5cm}
\begin{center}

{\LARGE \textbf {
On the Validity of the Effective Field Theory
\\
[0.01cm]
 for Dark Matter Searches at the LHC
 \\
 [0.3cm]
 Part II: Complete  Analysis for the $s$-channel
}}
\\ [1.5cm]

{\large
\textsc{Giorgio Busoni}$^{\rm a,}$\footnote{\texttt{giorgio.busoni@sissa.it}},
\textsc{Andrea De Simone}$^{\rm a, }$\footnote{\texttt{andrea.desimone@sissa.it}},
}
\\[0.2cm]
{\large
\textsc{Johanna Gramling}$^{\rm  b, }$\footnote{\texttt{johanna.gramling@unige.ch}},
\textsc{Enrico Morgante}$^{\rm b, }$\footnote{\texttt{enrico.morgante@unige.ch}},
\textsc{Antonio Riotto}$^{\rm b, }$\footnote{\texttt{antonio.riotto@unige.ch}}
}
\\[1cm]

\large{
$^{\rm a}$ 
\textit{SISSA and INFN, Sezione di Trieste, via Bonomea 265, I-34136 Trieste, Italy}\\
\vspace{1.5mm}
$^{b}$ 
\textit{Section de Physique, Universit\'e de Gen\`eve,\\
24 quai E. Ansermet, CH-1211 Geneva, Switzerland}
}
\end{center}

\vspace{0.5cm}

\begin{center}
\textbf{Abstract}
\begin{quote}
We generalize in several directions our  recent analysis of the limitations to the use of the effective field theory approach to study dark matter at the LHC. Firstly, we study the full list of operators connecting  fermion DM to quarks and gluons,  corresponding to integrating out a heavy mediator
in the $s$-channel; secondly,  we provide analytical results for the validity of the EFT description for both  $\sqrt{s}=8$ {\rm TeV} and $14$ {\rm TeV}; thirdly,
we make use of a  MonteCarlo event generator approach to assess the validity of our analytical conclusions. We apply our results to revisit the current collider  bounds on the ultraviolet cut-off scale of the effective field
theory and show that these bounds are weakened once
 the validity conditions of the  effective field theory are imposed.
\end{quote}
\end{center}

\def\thefootnote{\arabic{footnote}}
\setcounter{footnote}{0}
\pagestyle{empty}

\newpage
\pagestyle{plain}
\setcounter{page}{1}

%%%%%%%%%%%%%%%%%%%%%%%%%%%%%%%%%%%%%%%%%%
\section{Introduction}
%%%%%%%%%%%%%%%%%%%%%%%%%%%%%%%%%%%%%%%%%%
While there are many  cosmological and astrophysical evidences that our universe contains
a sizable amount of dark Matter (DM),  i.e.~a component which clusters at small scales, its nature is still a mystery. Various considerations point
towards the possibility that DM is made of neutral particles  whose mass and interactions  are dictated by  physics in the electroweak
energy range. If so,  the DM  relic density of these particles, assuming they were in thermal equilibrium during the evolution of the universe,  turns out to
be
\begin{equation}\label{Omega}
\left(\frac{\Omega_{\rm DM} h^2}{0.110}\right) \approx\frac{3 \times 10^{-26} {\rm cm}^3/{\rm sec}}{ \langle \sigma v \rangle_{\rm ann}},
\end{equation}
where  $\langle \sigma v \rangle_{\rm ann} $ is the (thermally-averaged)  annihilation cross
section.
A  weak interaction strength provides  the abundance in the right range measured by the Planck collaboration: $\Omega_{\rm DM}=0.315\pm0.0175$~\cite{Planck}.
This numerical coincidence
represents the main  reason why it is generically believed that DM is
made of weakly-interacting particles with a  mass in the range
$(10^2-10^4)$ GeV. 

Currently, there are several ways to search for such DM
candidates. Apart from the indirect \cite{review} and direct \cite{review direct} searches,  DM particles (if they are light enough) might reveal themselves in particle colliders, namely at the LHC.
Many LHC searches for DM  are based on the idea of looking at events with missing energy plus a single jet or photon, emitted from the initial state in $pp$ collisions (for alternative kinds of DM searches
at the LHC see e.g.~Refs.~\cite{DeSimone:2010tf, Bai:2011wy, Fox:2012ee, Cotta:2012nj, Bai:2012xg, Bell:2012rg})
\be
p p \to \chi+\overline{\chi}+{\rm jet/photon},
\ee
where $\chi$ indicates the DM particle. Several results are already available from two LHC collaborations \cite{monojetATLAS1, monojetCMS1, monojetATLAS2, monojetCMS2,
monogammaATLAS1, monogammaCMS1, monogammaATLAS2, monogammaCMS2}.

In order to avoid the overwhelming model-dependence introduced by the plethora of DM models discussed in the literature, DM searches at the LHC have made use of  the 
 Effective Field Theory (EFT) 
 \cite{Beltran:2010ww,goo,Bai:2010hh,Goodman:2010ku,Rajaraman:2011wf,fox,D,C,ds,
 Dreiner:2013vla, Chen:2013gya}. This  approach is a very powerful and economical way to
grasp the main features of a physical process, only in terms of the degrees of freedom
which are excited at the scale of the process. EFT techniques are successfully applied
in many branches of physics, and in particular they have become a standard way to present
experimental results for DM searches.

However, as far as collider searches are concerned, with the LHC being such a powerful machine,
it is not guaranteed that the events used to constrain an effective interaction are not occurring
at an energy scale larger than the cutoff scale of the effective description. In other words,
some (or many) events of DM production may occur with such a high momentum transfer that the EFT
is not a good description anymore.
The question about the validity of the EFT for collider searches of DM has become
pressing (see also Refs.~\cite{Busoni:2013lha, Buchmueller:2013dya, Goodman:2010ku, Dreiner:2012xm, Cotta:2012nj,  Fox:2012ru, An:2012va, Shoemaker:2011vi, Fox:2011fx}), especially in the perspective of analysing the data from the future LHC run at (13-14) {\rm TeV}.

Let us consider a simple model where there is a heavy mediator of mass $M$, 
to which the quarks and DM are coupled
with couplings $g_q$ and $g_\chi$, respectively.
The EFT is a good approximation only at low energies. Indeed, it is possible at low energies to integrate out the heavy mediator from the theory and obtain a tower of operators.
The matching condition of the ultra-violet (UV) theory with the mediator and its low-energy effective counterpart implies $\Lambda={M}/{\sqrt{g_q g_\chi}}$.
A DM production event occurs at an energy at which the EFT is reliable as long as $Q_{\rm tr}<M$, where $Q_{\rm tr}$ is the  momentum transfer in the process;
this, together with the condition of perturbativity of the couplings $g_{q,\chi}<4\pi$, 
implies
\be
\Lambda>\frac{Q_{\rm tr}}{\sqrt{g_qg_\chi}}>\frac{Q_{\rm tr}}{4\pi}\,.
\label{cond1}
\ee
If, in addition, one assumes the momentum transfer to occur in the $s$-channel, then
kinematics imposes $Q_{\rm tr}>2m_{\rm DM}$, so Eq.~(\ref{cond1}) becomes
\be
\Lambda>\frac{m_{\rm DM}}{2\pi}\,.
\label{mover2pi}
\ee
This is a very minimal requirement which is refined event-by-event by the stronger condition 
Eq.~(\ref{cond1}), which depends on $m_{\rm DM}$ through $Q_{\rm tr}$. 
It is clear
that the details of condition (\ref{cond1}) depend on the values of the couplings in the UV
theory.
In the following, for definiteness, we will mostly identify the mass of the new degrees of freedom $M$ with
the suppression scale of the operator $\Lambda$. This is equivalent to consider couplings in the 
UV theory of ${\cal O}(1)$. So, we will deal with the condition (but we will discuss also  the impact of taking couplings larger than 1) 
\be
Q_{\rm tr}\lesssim \Lambda\label{condition}\,.
\ee
In Ref. \cite{Busoni:2013lha} we have started the discussion of
the limitations to the use of the EFT  approach for DM searches at the LHC by  adopting a toy model
where the heavy mediator is exchanged in the $s$-channel and by 
introducing  a few quantities which quantify the error made when using effective operators to describe processes with very high momentum transfer. Our criteria indicated up to what cutoff energy scale, and with what precision, the effective description is valid, depending on the DM mass and couplings.
In this paper we significantly extend  our previous work along four different directions:
\begin{enumerate}
\item we consider the full list of operators connecting  fermion DM to quarks and corresponding to integrating out the heavy mediator
in the $s$-channel; 
\item we provide analytical results for the validity of the EFT description for both  $\sqrt{s}=8$ {\rm TeV} and $14$ {\rm TeV};
\item  we  follow a MonteCarlo approach to assess the validity of the EFT
and compare this fully numerical results with the analytical calculations;
\item we apply our results to revisit the current experimental bounds on the effective 
operator scale;  by requiring that only the events which are ``safe'' from the EFT point of view
should be considered, the bounds get weakened.
\end{enumerate}
The rest of the present paper is organized as follows.
In Section \ref{sec:validity} we present and discuss the results of our analytical approach
to assess the validity of EFT. In Section \ref{sec:numerical}, the fully numerical approach
is described and the results are compared with the analytical calculations. In Section \ref{sec:interp}
we analyze the impact of the limitation of the validity of the EFT for the current limits from the LHC searches. 
Finally, we draw our conclusions in 
Section \ref{sec:conclusions}. The details of the analytical results can be found in the Appendix \ref{app:crosssect}.

%%%%%%%%%%%%%%%%%%%%%%%%%%%%%%%%%%%%%%%%%%
\section{Validity of the EFT: analytical approach}
\label{sec:validity}
%%%%%%%%%%%%%%%%%%%%%%%%%%%%%%%%%%%%%%%%%%

\subsection{Operators and cross sections}
The starting point of our analysis is the list  of the 18 operators  reported in Tab.~\ref{table:operators} which are commonly used in the literature 
\cite{Beltran:2010ww}. 
We have considered not only the operators connecting the DM fermion to quarks (D1-D10), but also those involving gluon field strengths (D11-D14).
Furthermore, the operators can originate from heavy mediators exchange in the $s$-channel. 
For instance, 
 the D1' (D5) operators may be originated by the tree-level $s$-channel exchange of a very
heavy scalar (vector) boson $S$ ($V_\mu$), with lagrangians
\bea
\mathscr{L}_{\rm D1'}&\supset& \frac{1}{2}M^2 S^2
-g_q \bar q q S
-g_\chi \bar \chi \chi S\, ,
\label{D1UV}\\
\mathscr{L}_{\rm D5}&\supset& \frac{1}{2}M^2 V^{\mu}V_{\mu} -g_q \bar q \gamma^{\mu} q V_{\mu}-g_\chi \bar \chi \gamma^{\mu} \chi V_{\mu}\,.
\label{D5UV}
\eea
Notice the presence of the ``primed'' operators $D1'$--$D4'$, very similar to the ones often considered $D1$--$D4$, respectively, but with a different normalization, independent of the quark masses.
In fact, they may arise from integrating out heavy scalars which do not take a vacuum expectation value and therefore do
not give rise to quark masses.
\footnote{
A normalization proportional to the quark mass is common in many models motivated by flavour physics, but in general the coefficient $\Lambda^3$ at the denominator can have a different form.
For example, if the effective operators come from a Naturalness-motivated new physics theory like Supersymmetry or Composite Higgs Models, assuming a $U(2)^3$ flavour symmetry \cite{Barbieri:2011ci,Barbieri:2012uh} the normalization would be
\be
\lambda_{t,b} \frac{1}{\Lambda^2} \frac{m_q}{m_{t,b}}
\ee
where $\Lambda$ is an energy scale of the order some TeV related to the Electroweak Symmetry Breaking and $m_{t,b},\lambda_{t,b}$ are the mass and the Yukawa coupling with the Higgs of the top/bottom quark, depending on whether the quark $q$ is up-like or down-like.
In the present work, we will be agnostic about this point, and we'll keep both the primed and unprimed operators into account on the same footing as all others.
}

\begin{table}[t!]
\centering
\begin{tabular}{ | c || c | c |}
  \hline                        
  Name & Operator & Coefficient    \\   \hline\hline
  D1 & $\bar\chi  \chi\;\bar q  q $ & ${m_q}/{\Lambda^3}$ \\  \hline
  D1' & $\bar\chi  \chi\;\bar q q$ & ${1}/{\Lambda^2}$ \\ \hline
  D2 & $\bar\chi \gamma^5 \chi\;\bar q q $ & $i{m_q}/{\Lambda^3}$ \\ \hline
  D2' & $\bar\chi \gamma^5 \chi\;\bar q q $ & ${i}/{\Lambda^2}$ \\ \hline
  D3 & $\bar\chi \chi\;\bar q \gamma^5  q $&  $i{m_q}/{\Lambda^3}$ \\ \hline
  D3' & $ \bar\chi \chi\;\bar  q \gamma^5 q$ & ${i}/{\Lambda^2}$ \\ \hline
  D4 & $\bar\chi \gamma^5 \chi\; \bar q \gamma^5 q $ & ${m_q}/{\Lambda^3}$ \\ \hline
  D4' & $\bar \chi \gamma^5  \chi\; \bar  q \gamma^5 q $ & ${1}/{\Lambda^2}$ \\ \hline
  D5 & $\bar \chi \gamma_\mu \chi\; \bar q \gamma^\mu q$ & ${1}/{\Lambda^2}$ \\ \hline
  D6 & $\bar\chi \gamma_\mu \gamma^5 \chi\; \bar  q \gamma^\mu q $ & ${1}/{\Lambda^2}$ \\ \hline
  D7 & $\bar \chi \gamma_\mu  \chi\; \bar q \gamma^\mu\gamma^5  q$ & ${1}/{\Lambda^2}$ \\ \hline
  D8 & $\bar \chi \gamma_\mu \gamma^5 \chi\; \bar q \gamma^\mu \gamma^5 q $ & ${1}/{\Lambda^2}$
  \\ \hline  
  D9 & $\bar \chi \sigma_{\mu\nu} \chi\; \bar q \sigma^{\mu\nu} q $ & ${1}/{\Lambda^2}$ \\  \hline  
 D10 & $\bar \chi \sigma_{\mu\nu} \gamma^5\chi\; \bar q \sigma^{\mu\nu} q \;$ & ${i}/{\Lambda^2}$ 
 \\  \hline  
 D11 & $\bar \chi \chi\; G^{\mu\nu} G_{\mu\nu} \;$ & ${\alpha_s}/{4\Lambda^3}$ \\  \hline  
 D12 & $\bar \chi  \gamma^5\chi\; G^{\mu\nu} G_{\mu\nu}\;$ & ${i\alpha_s}/{4\Lambda^3}$ \\  \hline  
  D13 & $\bar \chi \chi\; G^{\mu\nu} \tilde{G}_{\mu\nu} \;$ & ${i\alpha_s}/{4\Lambda^3}$ \\  \hline  
 D14 & $\bar \chi  \gamma^5\chi\; G^{\mu\nu} \tilde G_{\mu\nu}\;$ & ${\alpha_s}/{4\Lambda^3}$ 
 \\  \hline  
\end{tabular}
\caption{\em Operators used throughout this work. The nomenclature is mostly taken from Ref.~\cite{Goodman:2010ku}.}
\label{table:operators}
\end{table}
We have computed the tree-level differential cross sections in the transverse momentum $p_{\rm T}$ and rapidity $\eta$ of the final jet for the hard scattering process with gluon radiation from the initial state
$f(p_1)+\bar f(p_2)\to \chi(p_3)+\chi(p_4)+g(k)$, where $f$ is either a quark (for operators D1-D10),
or a gluon (for operators D11-D14).
The results are conveniently written in terms of the momentum transfer  in the $s$-channel
\be
Q_{\rm tr}^2= (p_1+p_2-k)^2=
{x_1 x_2 s}-\sqrt{s}\,p_{\rm T}\left(x_1e^{-\eta}+x_2e^{\eta}\right)\,,
\label{Qtransfer}
\ee
where $x_1, x_2$ are the fractions of momentum carried by initial partons and $\eta, p_{\rm T}$ are the pseudo-rapidity and the transverse momentum of the final state gluon, respectively.
The expressions are of course valid for all admitted values of the parameters. It’s only when integrated numerically over the PDFs and over $\eta, p_{\rm T}$ that the dependence on these values comes in.
We obtain
\bea
\left.\dfrac{\de^2\hat\sigma}{\de p_{\rm T}\de\eta}\right\vert_{D1'}&=&
\frac{ \alpha_s}{36 \pi^2}
\frac{1}{p_{\rm T}}\frac{1}{\Lambda^4}
\frac{\left[Q_{\rm tr}^2-4m_{\rm DM}^2\right]^{3/2}\left[1+\frac{Q_{\rm tr}^4}{(x_1 x_2 s)^2}\right]}{\,Q_{\rm tr}}
\label{d2sigmaefflab1}\,, \\
\left.\dfrac{\de^2\hat\sigma}{\de p_{\rm T}\de\eta}\right\vert_{D4'}&=&
\frac{ \alpha_s}{36 \pi^2}
\frac{1}{p_{\rm T}}\frac{1}{\Lambda^4}
Q_{\rm tr}\left[Q_{\rm tr}^2-4m_{\rm DM}^2\right]^{1/2}\left[1+\frac{Q_{\rm tr}^4}{(x_1 x_2 s)^2}\right]
\label{d2sigmaefflab4}\,,
\eea
\bea
\left.\dfrac{\de^2\hat\sigma}{\de p_{\rm T}\de\eta}\right\vert_{D5}&=&
\frac{ \alpha_s}{27 \pi^2}
\frac{1}{p_{\rm T}}\frac{1}{\Lambda^4}
\frac{\left[Q_{\rm tr}^2-4m_{\rm DM}^2\right]^{1/2}\left[Q_{\rm tr}^2+2m_{\rm DM}^2\right]
\left[1+\frac{Q_{\rm tr}^4}{(x_1 x_2 s)^2}-2\frac{p_{\rm T}^2}{x_1 x_2 s}\right]}{\,Q_{\rm tr}}
\label{d2sigmaefflab5}\,, \\
\left.\dfrac{\de^2\hat\sigma}{\de p_{\rm T}\de\eta}\right\vert_{D8}&=&
\frac{ \alpha_s}{27 \pi^2}
\frac{1}{p_{\rm T}}\frac{1}{\Lambda^4}
\frac{\left[Q_{\rm tr}^2-4m_{\rm DM}^2\right]^{3/2}\left[1+\frac{Q_{\rm tr}^4}{(x_1 x_2 s)^2}-2\frac{p_{\rm T}^2}{x_1 x_2 s}\right]}{\,Q_{\rm tr}}
\label{d2sigmaefflab8}\,,\\ 
\left.\dfrac{\de^2\hat\sigma}{\de p_{\rm T}\de\eta}\right\vert_{D9}&=& 
\frac{2 \alpha_s}{27 \pi^2}
\frac{1}{p_{\rm T}}\frac{1}{\Lambda^4}
\frac{\sqrt{Q_{\rm tr}-4m_{\rm DM}^2}\left[Q_{\rm tr}^2+2m_{\rm DM}^2\right]
\left[1+\frac{Q_{\rm tr}^4}{( x_1 x_2 s)^2}+4p_{\rm T}^2\left(\frac{1}{Q_{\rm tr}^2}-\frac{1}{x_1 x_2 s}\right)\right]}{\,Q_{\rm tr}}\,,\nn\\ &&
\label{d2sigmaefflab9}\\
\left.\dfrac{\de^2\hat\sigma}{\de p_{\rm T}\de\eta}\right\vert_{D11}&=&
\frac{3\alpha_s^3}{256\pi^2\Lambda^6}\frac{(x_1x_2s)^3}{(Q_{\rm tr}^2-x_1x_2s)^2}\frac{(Q_{\rm tr}^2-4 m_{\rm DM}^2)^{3/2}}{p_{\rm T}Q_{\rm tr}}
\left[1-4\frac{Q_{\rm tr}^2-p_{\rm T}^2}{ x_1 x_2 s}+\frac{8Q_{\rm tr}^4+21p_{\rm T}^4}{(x_1 x_2 s)^2}\right.\nn\\
&&
-2Q_{\rm tr}^2\frac{5Q_{\rm tr}^4+4Q_{\rm tr}^2p_{\rm T}^2+5p_{\rm T}^4}{(x_1 x_2 s)^3}+Q_{\rm tr}^4\frac{8Q_{\rm tr}^4+8Q_{\rm tr}^2p_{\rm T}^2+5p_{\rm T}^4}{(x_1 x_2 s)^4}-4Q_{\rm tr}^8\frac{Q_{\rm tr}^2+p_{\rm T}^2}{ (x_1 x_2 s)^5}\nn\\
&&\left. +\frac{Q_{\rm tr}^{12}}{(x_1 x_2 s)^6}\right]\,, \label{d2sigmaefflab11} \\
\left.\dfrac{\de^2\hat\sigma}{\de p_{\rm T}\de\eta}\right\vert_{D12}&=&
\frac{3\alpha_s^3}{256\pi^2\Lambda^6}\frac{(x_1x_2s)^3}{(Q_{\rm tr}^2-x_1x_2s)^2} \frac{Q_{\rm tr}\sqrt{Q_{\rm tr}^2-4 m_{\rm DM}^2}}{p_{\rm T}}
\left[1-4\frac{Q_{\rm tr}^2-p_{\rm T}^2}{ x_1 x_2 s}+\frac{8Q_{\rm tr}^4+21p_{\rm T}^4}{(x_1 x_2 s)^2}
\right.\nn\\
&&
-2Q_{\rm tr}^2\frac{5Q_{\rm tr}^4+4Q_{\rm tr}^2p_{\rm T}^2+5p_{\rm T}^4}{(x_1 x_2 s)^3}+ Q_{\rm tr}^4\frac{8Q_{\rm tr}^4+8Q_{\rm tr}^2p_{\rm T}^2+5p_{\rm T}^4}{(x_1 x_2 s)^4}-4Q_{\rm tr}^8\frac{Q_{\rm tr}^2+p_{\rm T}^2}{ (x_1 x_2 s)^5}\nn\\
&&\left. +\frac{Q_{\rm tr}^{12}}{(x_1 x_2 s)^6}\right]\,,\label{d2sigmaefflab12}
\eea 
\bea
\left.\dfrac{\de^2\hat\sigma}{\de p_{\rm T}\de\eta}\right\vert_{D13}&=&
\frac{3\alpha_s^3}{256\pi^2\Lambda^6}\frac{(x_1x_2s)^3}{(Q_{\rm tr}^2-x_1x_2s)^2}\frac{(Q_{\rm tr}^2-4 m_{\rm DM}^2)^{3/2}}{p_{\rm T}Q_{\rm tr}}
\left[1-4\frac{Q_{\rm tr}^2}{ x_1 x_2 s}+\frac{8Q_{\rm tr}^4+8Q_{\rm tr}^2p_{\rm T}^2+5p_{\rm T}^4}{(x_1 x_2 s)^2}\right.\nn\\
&&
 -2Q_{\rm tr}^2\frac{5Q_{\rm tr}^4+6Q_{\rm tr}^2p_{\rm T}^2-3p_{\rm T}^4}{(x_1 x_2 s)^3}+Q_{\rm tr}^4\frac{8Q_{\rm tr}^4+8Q_{\rm tr}^2p_{\rm T}^2+5p_{\rm T}^4}{(x_1 x_2 s)^4}-4Q_{\rm tr}^8\frac{Q_{\rm tr}^2+p_{\rm T}^2}{ (x_1 x_2 s)^5}\nn\\
&&\left. +\frac{Q_{\rm tr}^{12}}{(x_1 x_2 s)^6}\right]\,,\label{d2sigmaefflab13}\\
\left.\dfrac{\de^2\hat\sigma}{\de p_{\rm T}\de\eta}\right\vert_{D14}&=&
\frac{3\alpha_s^3}{256\pi^2\Lambda^6}\frac{(x_1x_2s)^3}{(Q_{\rm tr}^2-x_1x_2s)^2} \frac{Q_{\rm tr}\sqrt{Q_{\rm tr}^2-4 m_{\rm DM}^2}}{p_{\rm T}}
\left[1-4\frac{Q_{\rm tr}^2}{ x_1 x_2 s}+\frac{8Q_{\rm tr}^4+8Q_{\rm tr}^2p_{\rm T}^2+5p_{\rm T}^4}{(x_1 x_2 s)^2}\right.\nn\\
&&
 -2Q_{\rm tr}^2\frac{5Q_{\rm tr}^4+6Q_{\rm tr}^2p_{\rm T}^2-3p_{\rm T}^4}{(x_1 x_2 s)^3}+Q_{\rm tr}^4\frac{8Q_{\rm tr}^4+8Q_{\rm tr}^2p_{\rm T}^2+5p_{\rm T}^4}{(x_1 x_2 s)^4}-4Q_{\rm tr}^8\frac{Q_{\rm tr}^2+p_{\rm T}^2}{ (x_1 x_2 s)^5}\nn\\
&&\left. +\frac{Q_{\rm tr}^{12}}{(x_1 x_2 s)^6}\right]\,.\label{d2sigmaefflab14}
\eea
The reader can find the details of the derivation of  Eqs.~(\ref{d2sigmaefflab1})-(\ref{d2sigmaefflab14})  
 in Appendix \ref{app:crosssect}. 
As for the other operators, we get 
\bea
\left.\dfrac{\de^2\hat\sigma}{\de p_{\rm T}\de\eta}\right\vert_{D2'}=
\left.\dfrac{\de^2\hat\sigma}{\de p_{\rm T}\de\eta}\right\vert_{D4'}\qquad
&\left.\dfrac{\de^2\hat\sigma}{\de p_{\rm T}\de\eta}\right\vert_{D3'}=
\left.\dfrac{\de^2\hat\sigma}{\de p_{\rm T}\de\eta}\right\vert_{D1'}&\qquad
\left.\dfrac{\de^2\hat\sigma}{\de p_{\rm T}\de\eta}\right\vert_{D6}=
\left.\dfrac{\de^2\hat\sigma}{\de p_{\rm T}\de\eta}\right\vert_{D8}\\
\left.\dfrac{\de^2\hat\sigma}{\de p_{\rm T}\de\eta}\right\vert_{D7}=
\left.\dfrac{\de^2\hat\sigma}{\de p_{\rm T}\de\eta}\right\vert_{D5}&\qquad&
\left.\dfrac{\de^2\hat\sigma}{\de p_{\rm T}\de\eta}\right\vert_{D9}=
\left.\dfrac{\de^2\hat\sigma}{\de p_{\rm T}\de\eta}\right\vert_{D10}\,,
\eea
in the limit of massless light quarks.
The operators $D1$--$D4$ are simply related to $D1'$--$D4'$ by a straightforward rescaling 
\be
\left.\dfrac{\de^2\hat\sigma}{\de p_{\rm T}\de\eta}\right\vert_{D1, D2, D3, D4}=
\left(\frac{m_q}{\Lambda}\right)^2
\left.\dfrac{\de^2\hat\sigma}{\de p_{\rm T}\de\eta}\right\vert_{{D1', D2', D3', D4'}}\,.
\label{primeunprime}
\ee
We checked that the differences between the cross sections for $D1'$--$D4'$ computed for $m_q\neq 0$  and those reported above assuming $m_q=0$ are at the per-mille level,
so the approximation $m_q = 0$ which we used in all our analytical calculations is justified.
The cross sections for the   UV completions of dim-6 operators, with $s$-channel exchange of a mediator
of mass $M_{\rm med}$, are simply obtained by the replacement $1/\Lambda^4 \to g_q^2 g_\chi^2/[Q_{\rm tr}^2-M_{\rm med}^2]^2$.

In order to get the cross sections initiated by the colliding protons one needs to average
over the PDFs. For example, for processes with initial state quarks
\be
\left.\dfrac{\de^2\sigma}{\de p_{\rm T}\de\eta}\right\vert_{Di}=
\sum_q\int\de x_1\de x_2 
[f_q(x_1)f_{\bar q}(x_2)+f_q(x_2)f_{\bar q}(x_1)] 
\left.\dfrac{\de^2\hat\sigma}{\de p_{\rm T}\de\eta}\right\vert_{Di}\,.
\ee
We have performed the analytical calculation only for the  emission of an initial state gluon (identified with the final jet observed experimentally). The extension to include also the smaller contribution coming
from initial radiation of quarks ($qg\to\chi\chi+q$) is done numerically in Section \ref{sec:numerical}.

\subsection{Results and discussion}

In what regions of the parameter space $(\Lambda, m_{\rm DM})$ is the effective description
accurate and reliable?
The truncation to the lowest-dimensional operator of the EFT expansion is  accurate only if the momentum transfer is smaller than an energy scale of the order of  $\Lambda$, see Eqs.~(\ref{condition}). 
Therefore we want to compute the fraction of events with momentum transfer 
lower than the EFT cutoff scale.
To this end we define the ratio of  the  cross section
obtained in the EFT with the requirement $Q_{\rm tr}<\Lambda$ on the
PDF integration domain, over the total cross section obtained in the EFT.
\be
R_\Lambda^{\rm tot}\equiv\frac{\sigma\vert_{Q_{\rm tr}<\Lambda}}{\sigma}
=\frac{\int_{p_{\rm T}^{\rm min}}^{p_{\rm T}^{\rm max}}\de p_{\rm T}\int_{-2}^2\de \eta
\left.\dfrac{\de^2\sigma}{\de p_{\rm T}\de\eta}\right\vert_{Q_{\rm tr}<\Lambda}}
{\int_{p_{\rm T}^{\rm min}}^{p_{\rm T}^{\rm max}}\de p_{\rm T}\int_{-2}^2\de \eta
\dfrac{\de^2\sigma}{\de p_{\rm T}\de\eta}}.
\label{ratiolambdatot}
\ee
To sum over the possible $p_{\rm T}, \eta$ of the jets, we integrate the differential cross sections
over values typically considered in the experimental searches.
We consider $p_{\rm T}^{\rm min}=500 \GeV$ (as used in the signal region SR4 of \cite{monojetATLAS2}), $|\eta|<2$ and the two cases with center-of-mass energies $\sqrt{s}=$ 8 {\rm TeV} and 14 {\rm TeV}.
 For $p_{\rm T}^{\rm max}$ we used 1, 2 {\rm TeV} for $\sqrt{s}=8, 14$ {\rm TeV}, respectively.
The sum over quark flavours is performed only considering $u,d,c,s$ quarks.

\begin{figure}[t!]
\centering
\includegraphics[width=0.45\textwidth]{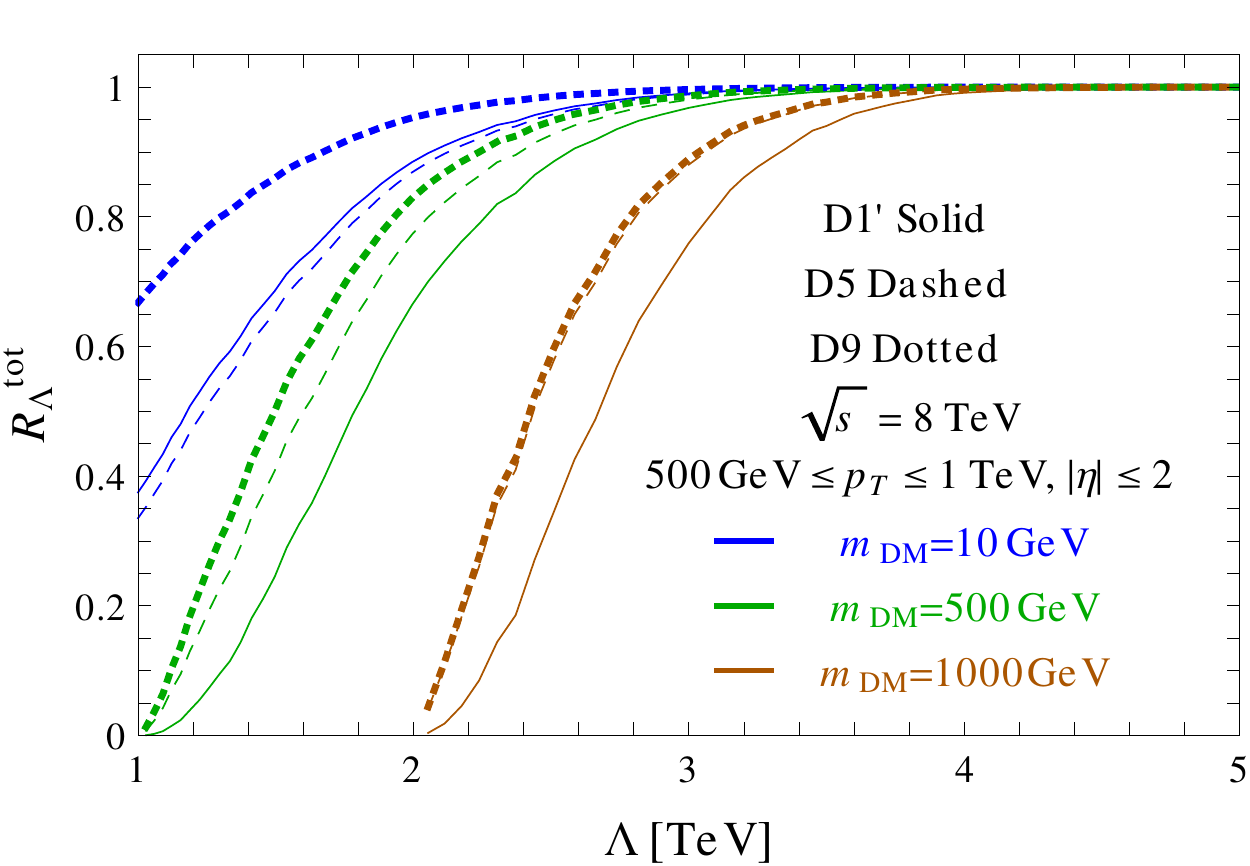}
\hspace{0.5cm}
\includegraphics[width=0.45\textwidth]{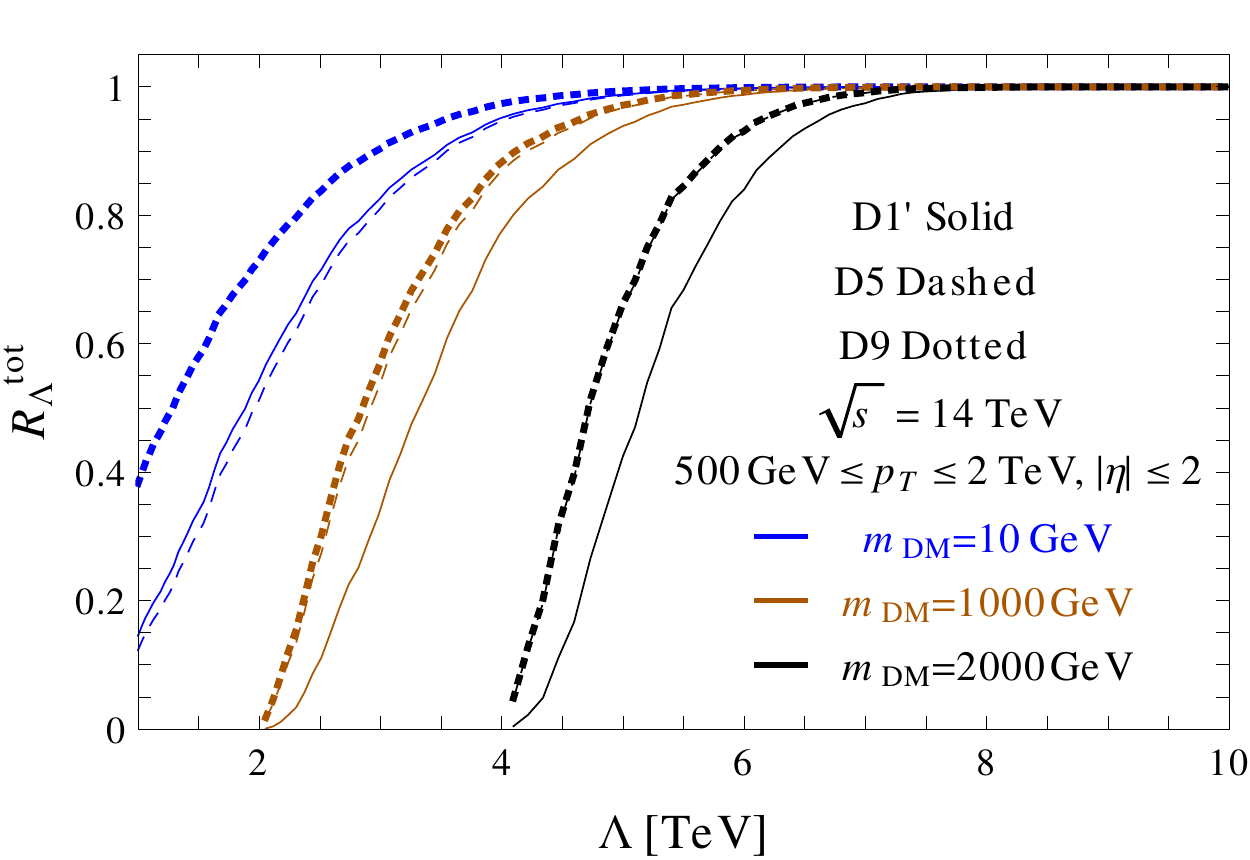}\\
\includegraphics[width=0.45\textwidth]{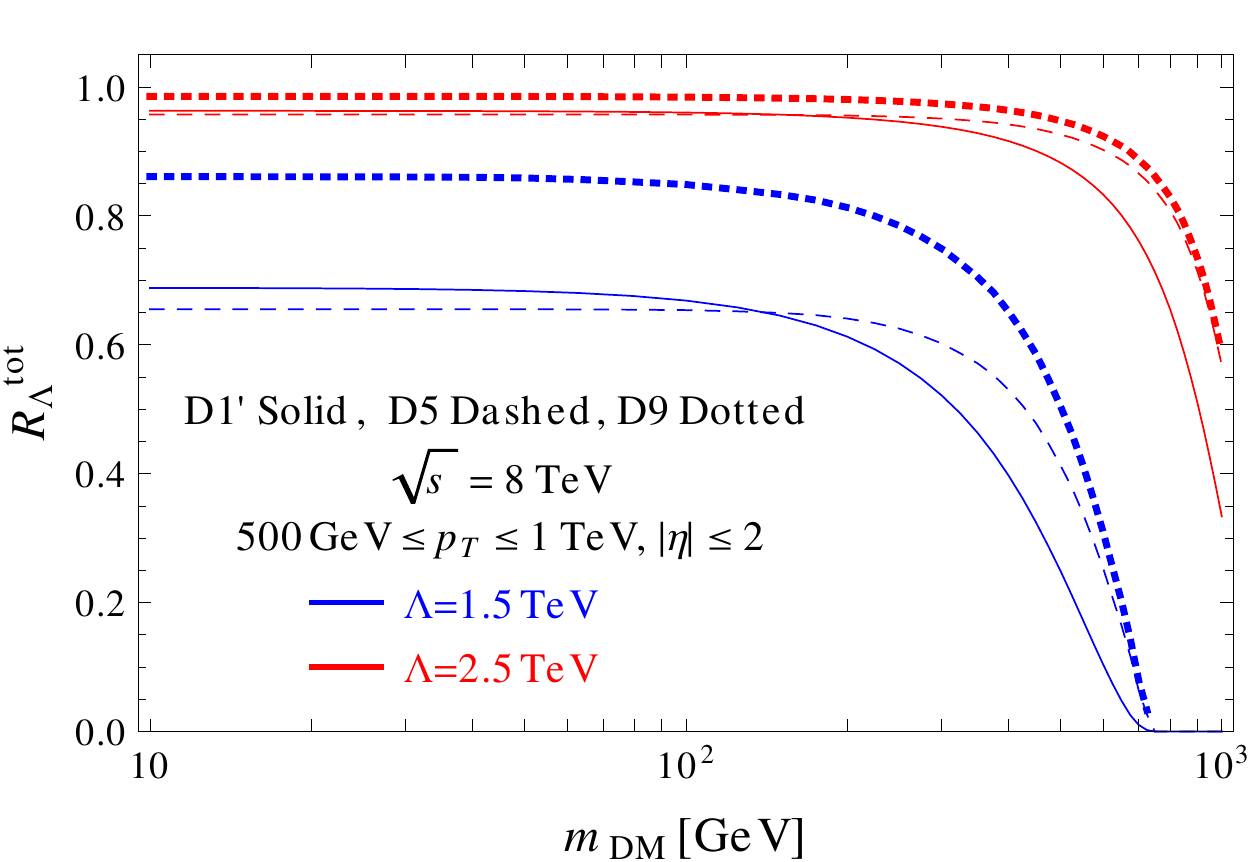}
\hspace{0.5cm}
\includegraphics[width=0.45\textwidth]{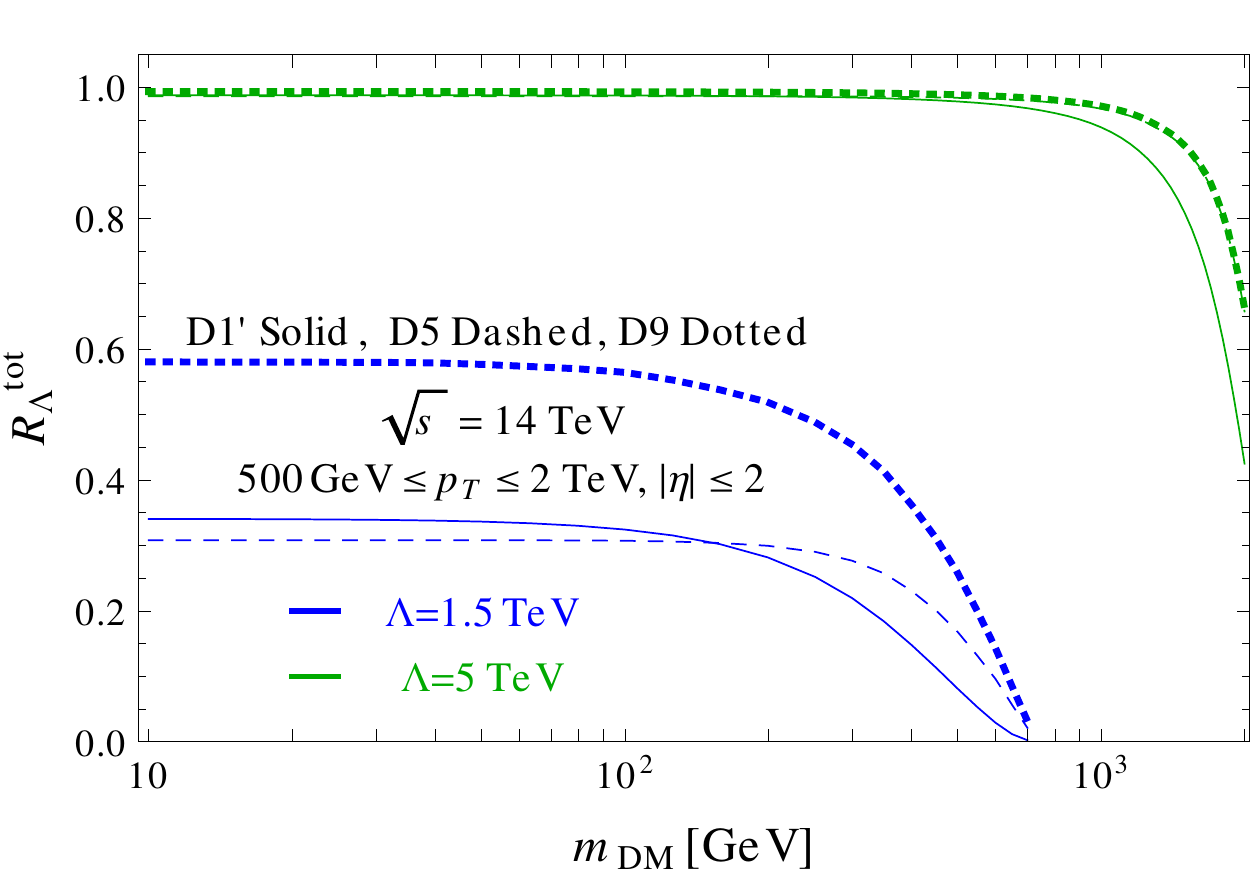}
\caption{ \em
 The ratio $R_{\Lambda}^{\rm tot}$ defined in Eq.~(\ref{ratiolambdatot}) for operators $D1'$ (solid lines),
  $D5$ (dashed lines) and $D9$ (dotted lines) as a function of $\Lambda$ and $m_{\rm DM}$,
  for $\sqrt{s}=8$ TeV (left panel) and $14$ TeV (right panel).
}
\label{fig:RLambdatot}
\end{figure}

We first study the behavior of the ratio $R_\Lambda^{\rm tot}$, as a function of $\Lambda$ and 
$m_{\rm DM}$ The results are shown in Fig.~\ref{fig:RLambdatot}. 
We show only results for representative operators $D1', D5, D9$.
This ratio $R_\Lambda^{\rm tot}$ gets closer to unity for large values of $\Lambda$, as in this case the effect of the cutoff becomes negligible. The ratio drops for large $m_{\rm DM}$ because the momentum
transfer increases in this regime.
This confirms our precedent  analysis of Ref.~\cite{Busoni:2013lha}, that
the EFT works better for large $\Lambda$ and small $m_{\rm DM}$.
Notice also that, going from $\sqrt{s}=8 {\rm TeV}$ to $\sqrt{s}=14 {\rm TeV}$, the results scale almost linearly with the energy, so for the same value of the ratio $m_{\rm DM}/\Lambda$ one obtains nearly the 
same $R_\Lambda^{\rm tot}$.

\begin{figure}[p!]
\centering
\includegraphics[width=0.45\textwidth]{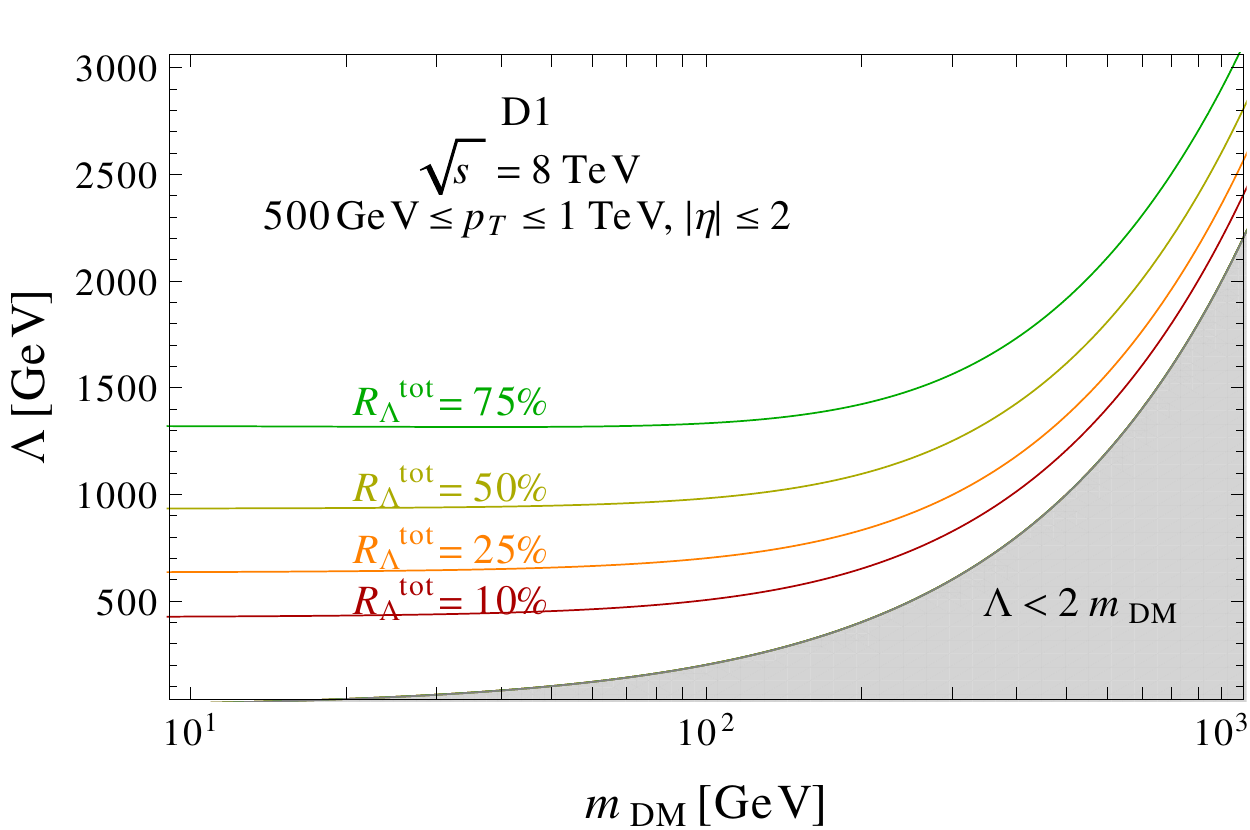} 
\hspace{0.5cm} 
\includegraphics[width=0.45\textwidth]{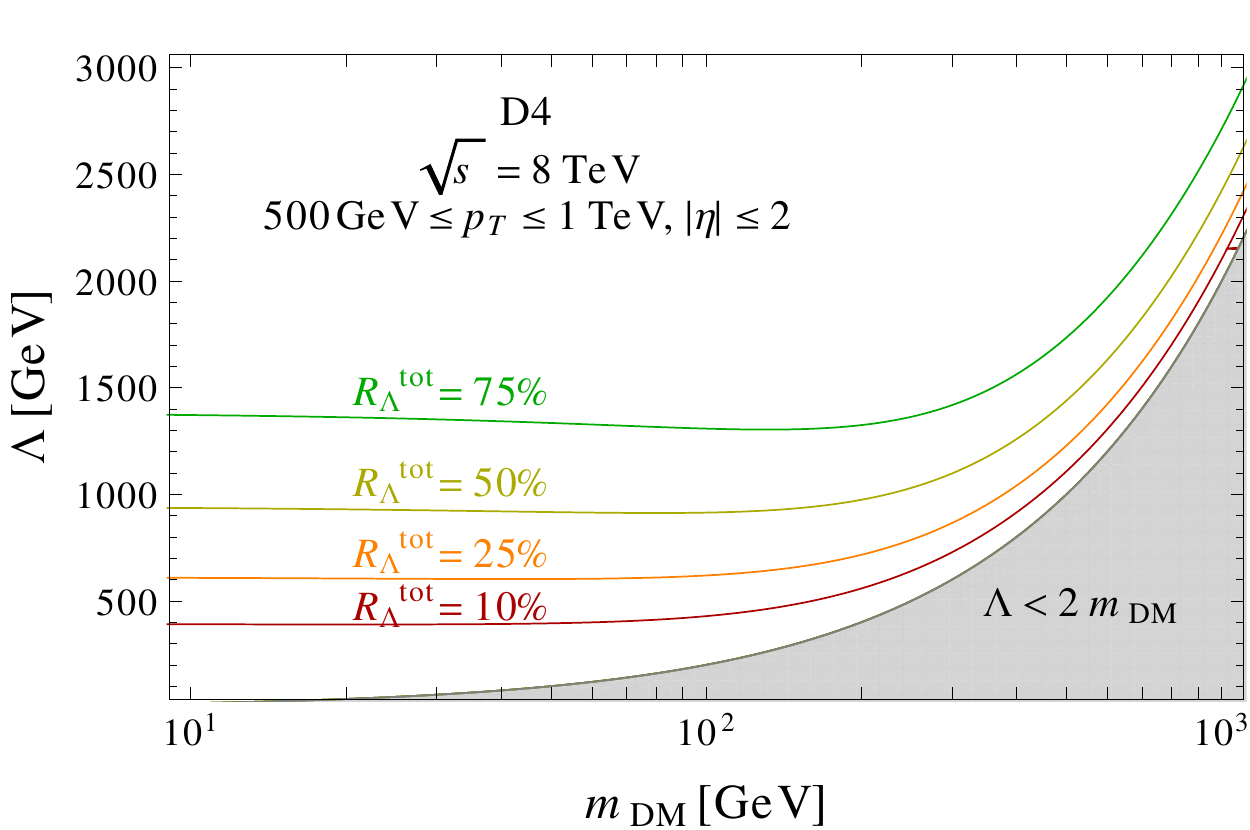}\\
\includegraphics[width=0.45\textwidth]{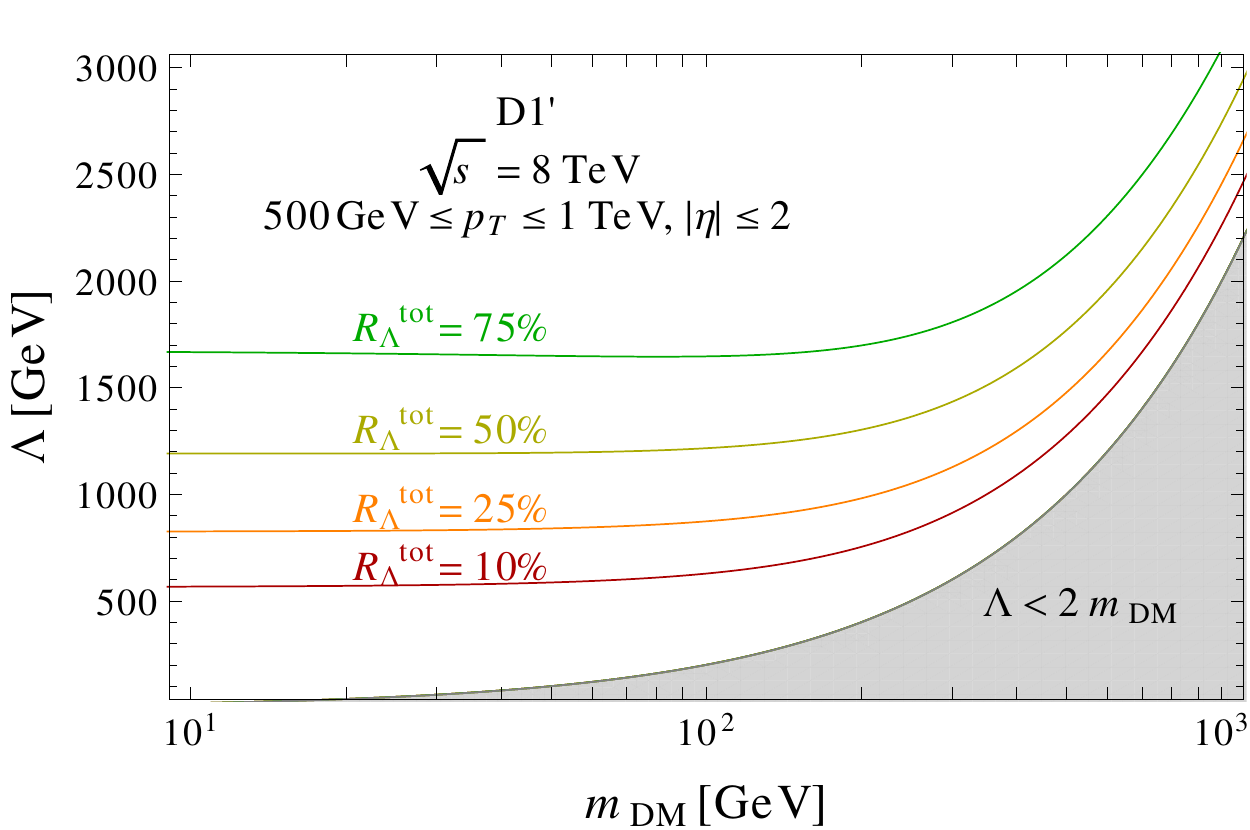}
\hspace{0.5cm}
\includegraphics[width=0.45\textwidth]{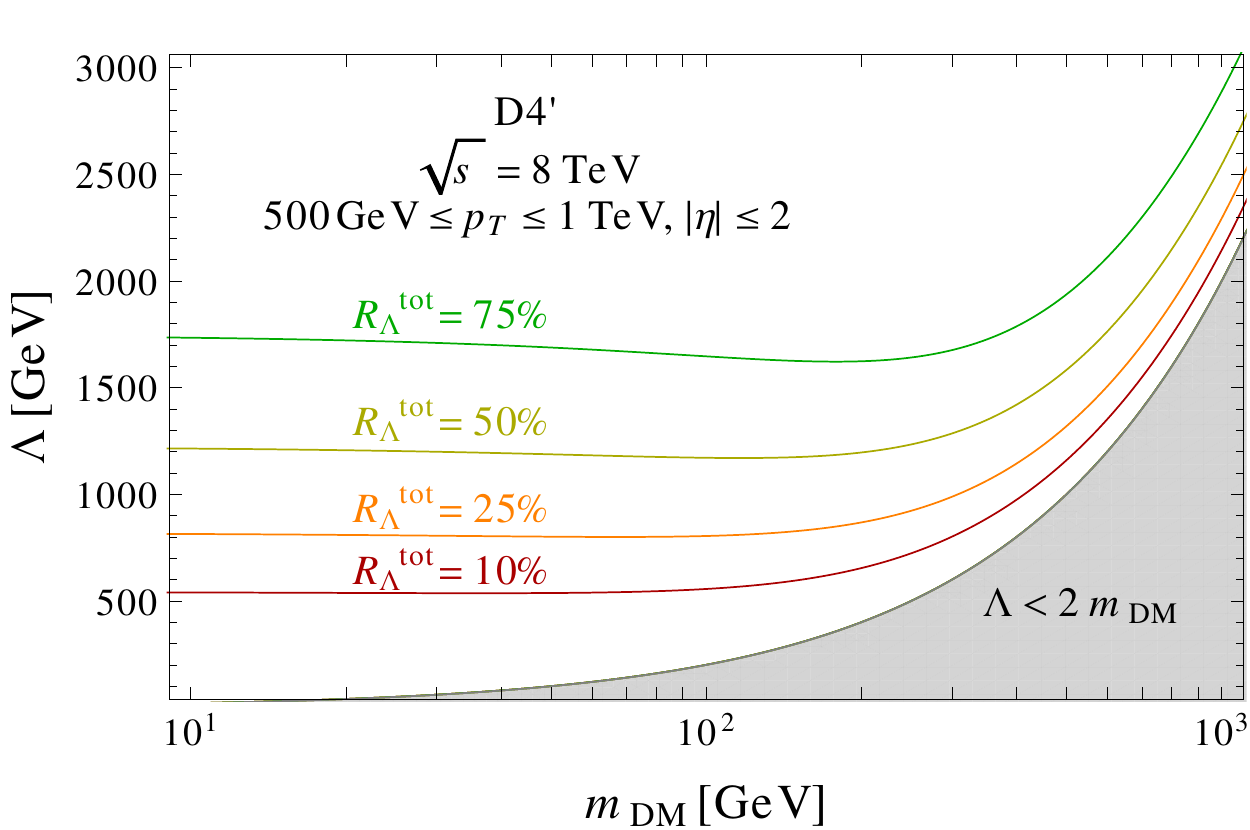}\\
\includegraphics[width=0.45\textwidth]{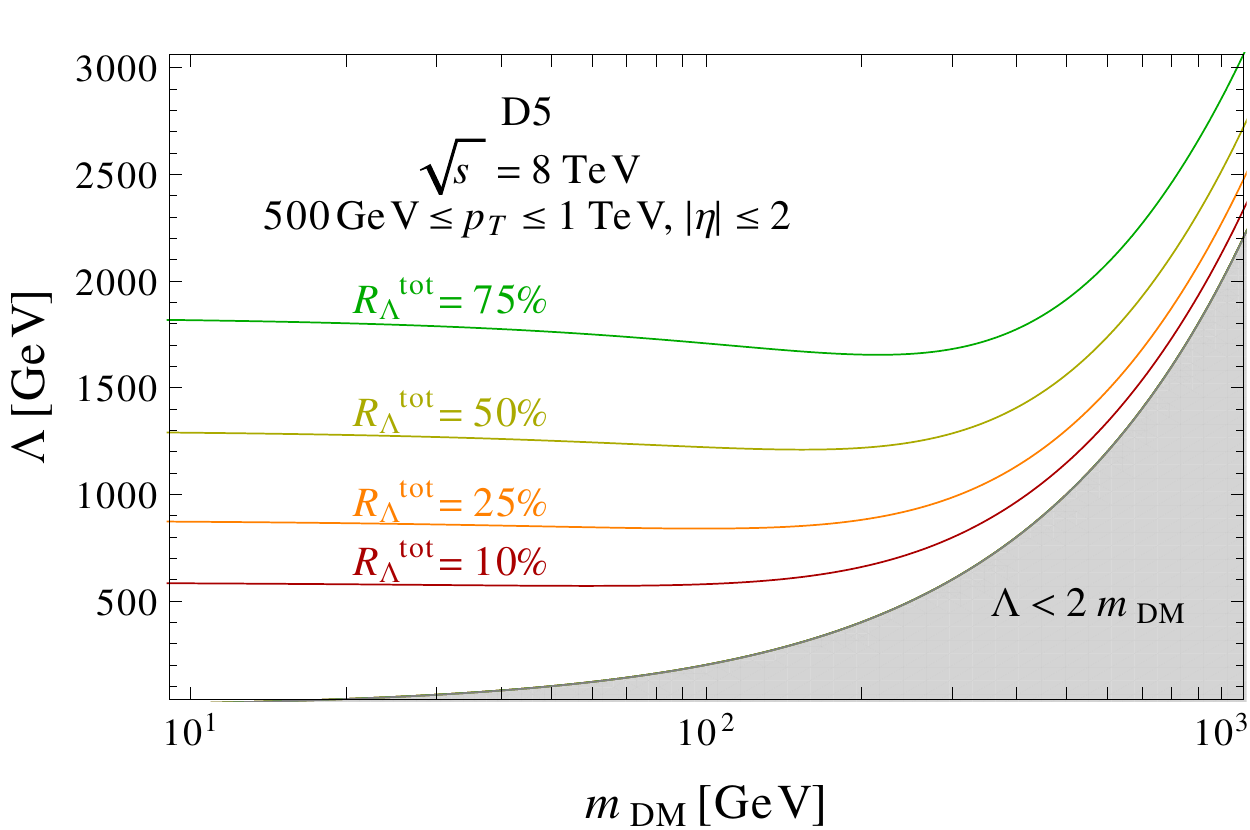}
\hspace{0.5cm}
\includegraphics[width=0.45\textwidth]{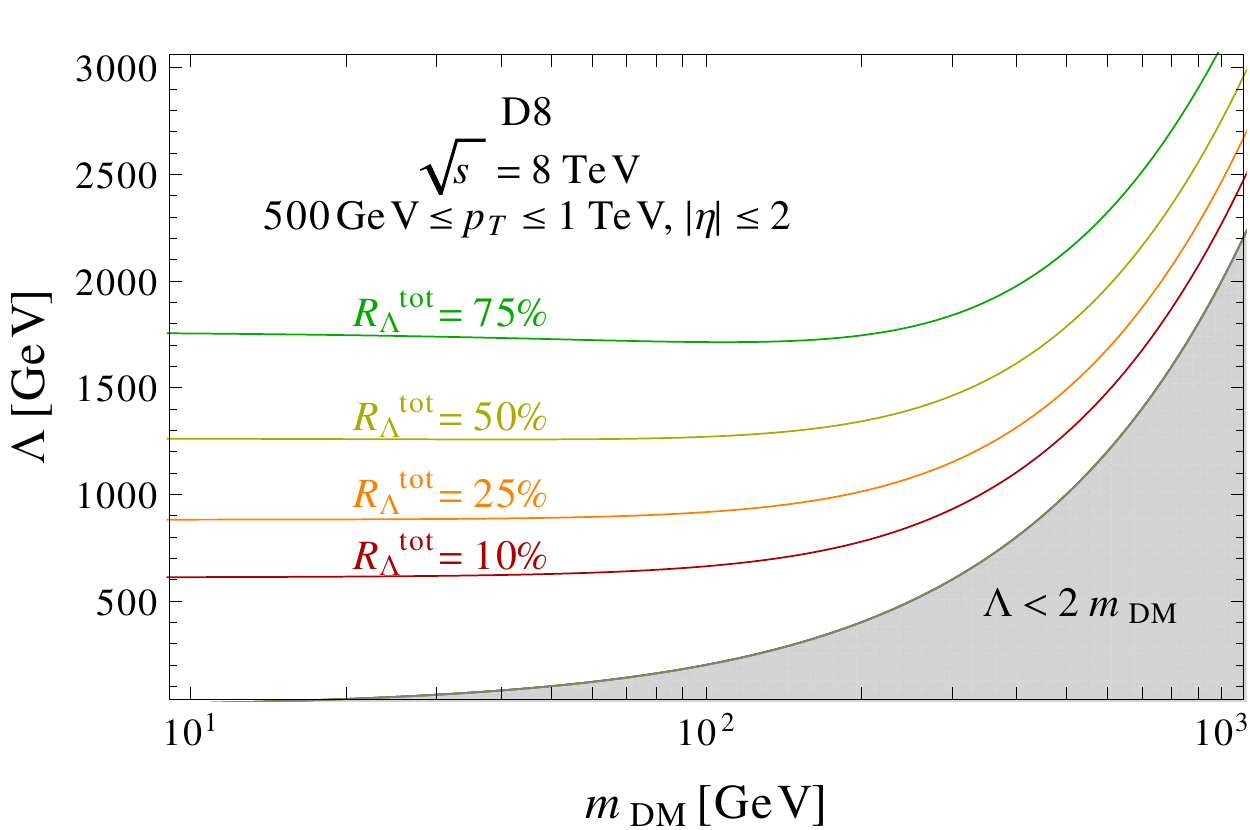}\\
\includegraphics[width=0.45\textwidth]{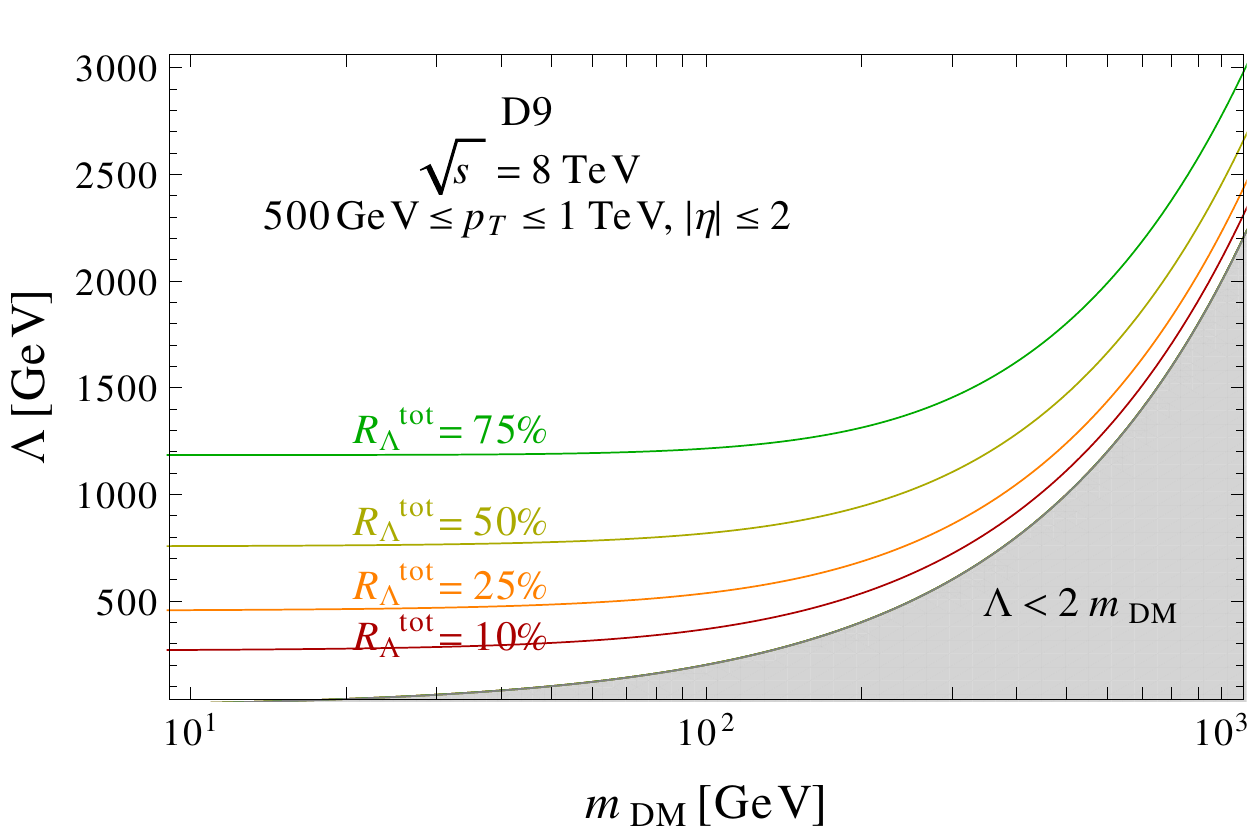}
\hspace{0.5cm}
\includegraphics[width=0.45\textwidth]{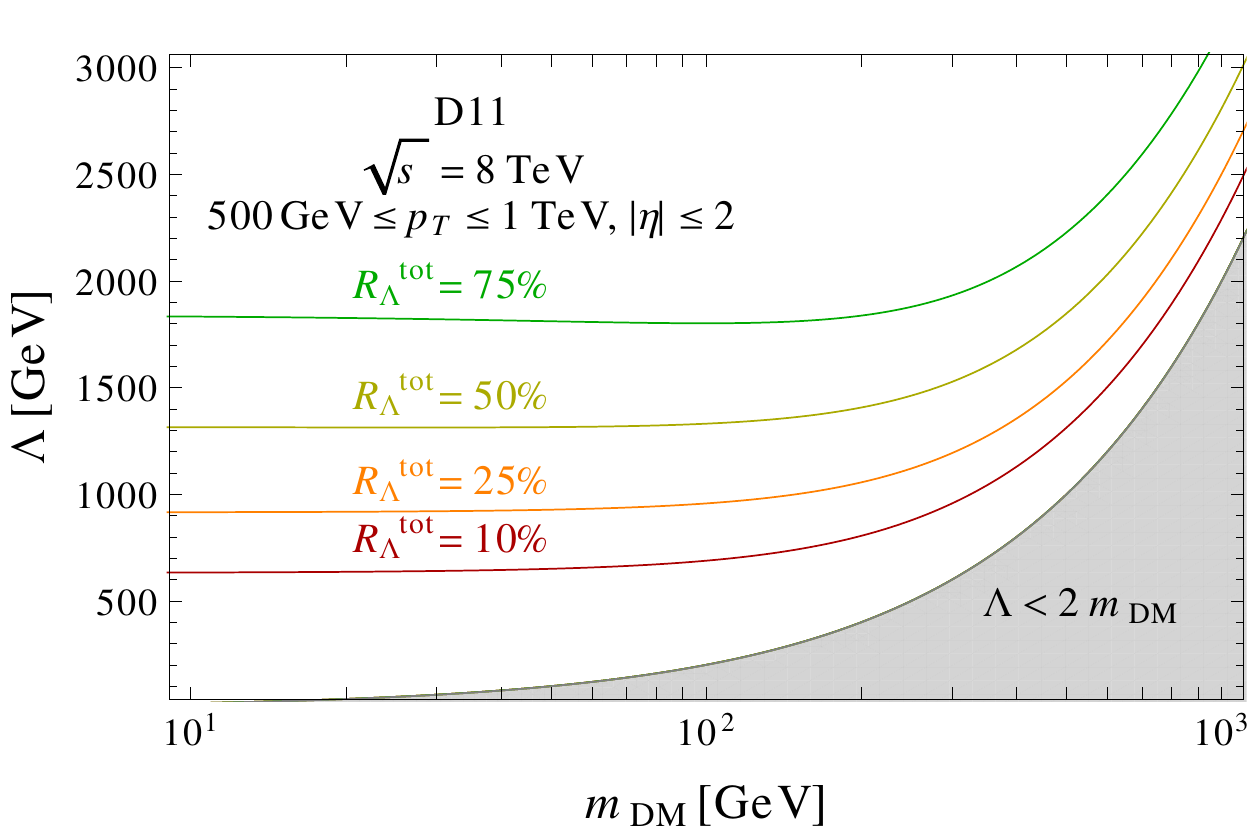}\\
\caption{ \em
Contours for the ratio $R_{\Lambda}^{\rm tot}$, defined in Eq.~(\ref{ratiolambdatot}), on the plane
$(m_{\rm DM}, \Lambda)$, for the different operators. We set $\sqrt{s}=8 {\rm TeV},
|\eta|\leq 2$ and $500 \GeV<p_{\rm T}<1 \TeV$. 
}
\label{fig:RLambdacontours}
\end{figure}

\begin{figure}[p!]
\centering
\includegraphics[width=0.45\textwidth]{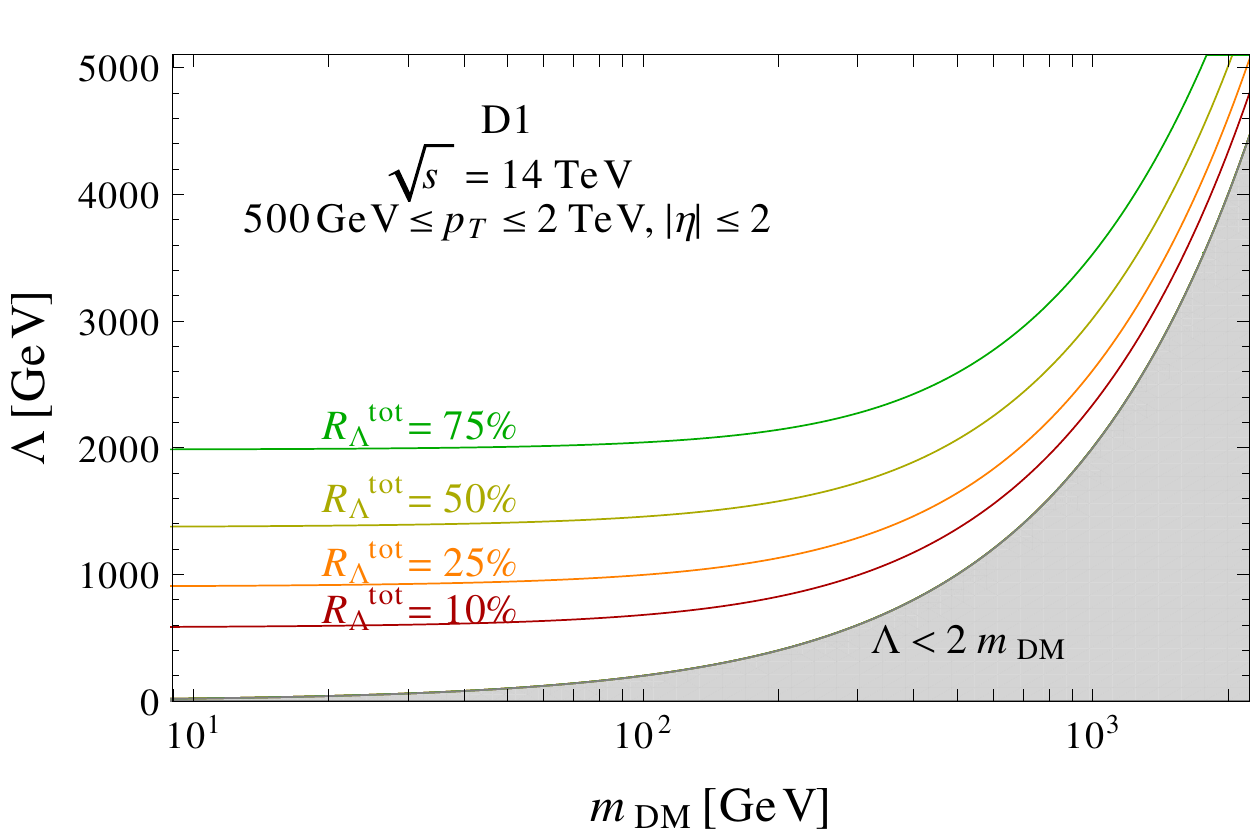} 
\hspace{0.5cm} 
\includegraphics[width=0.45\textwidth]{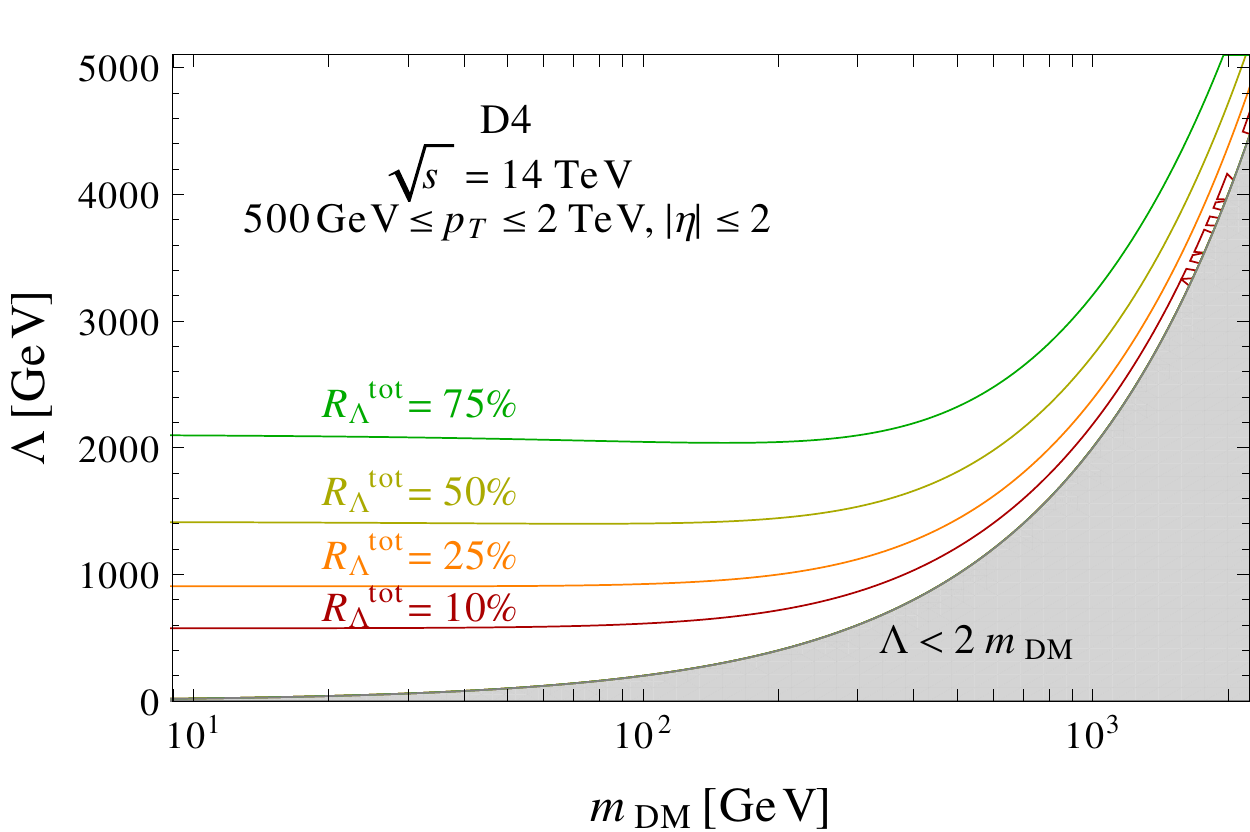}\\
\includegraphics[width=0.45\textwidth]{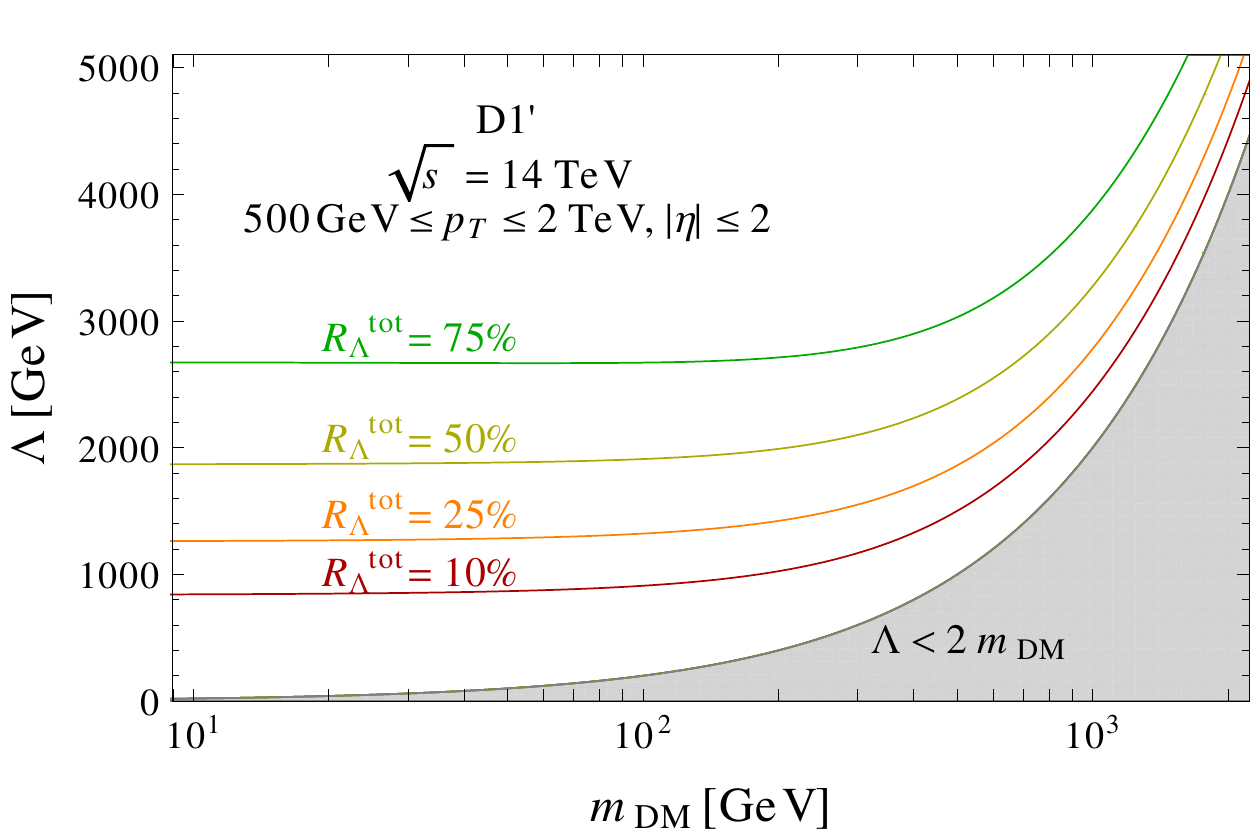}
\hspace{0.5cm}
\includegraphics[width=0.45\textwidth]{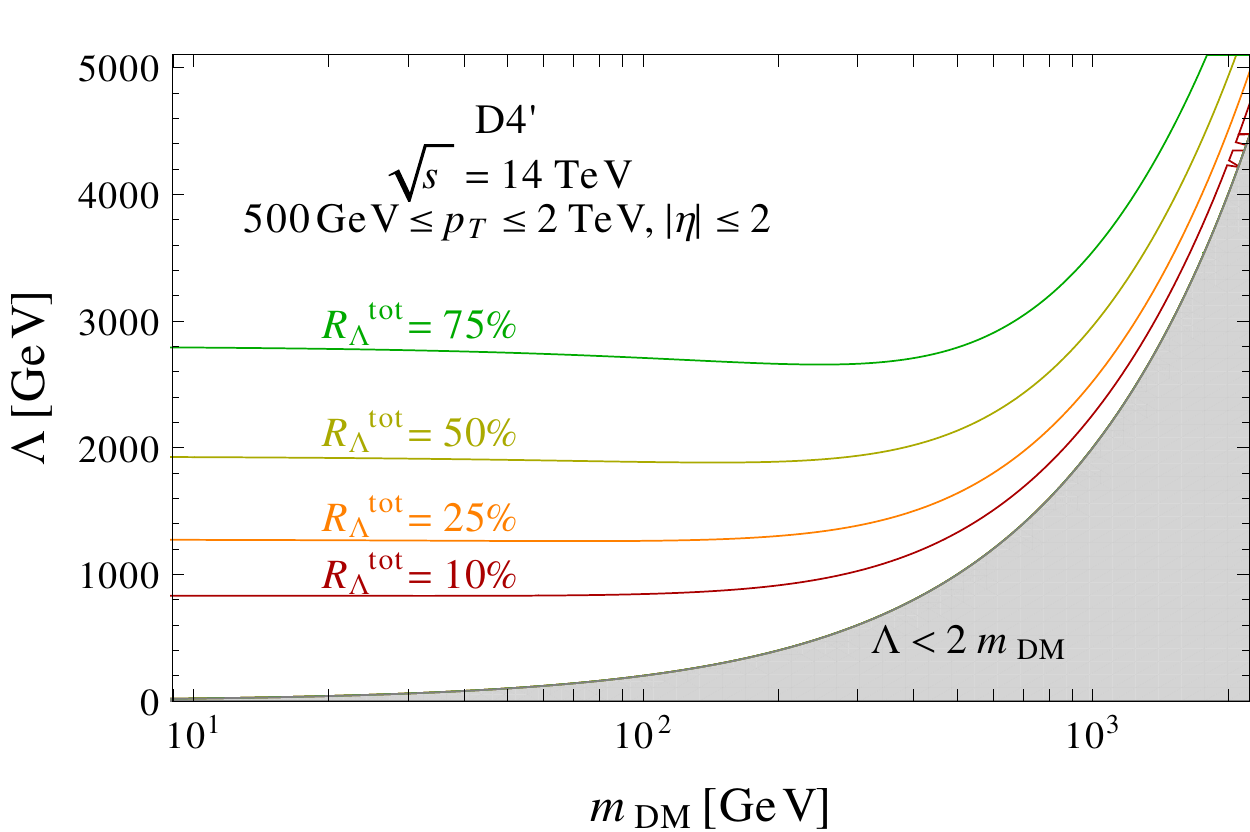}\\
\includegraphics[width=0.45\textwidth]{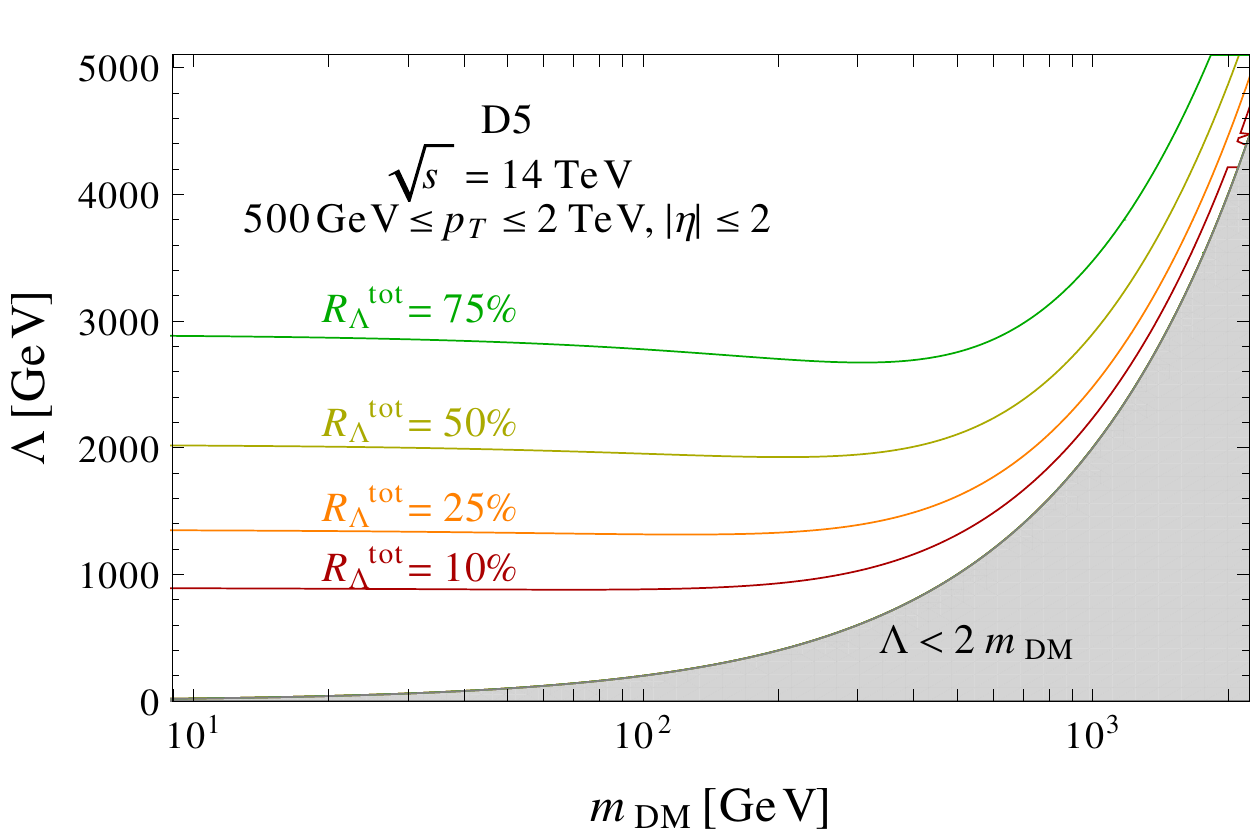}
\hspace{0.5cm}
\includegraphics[width=0.45\textwidth]{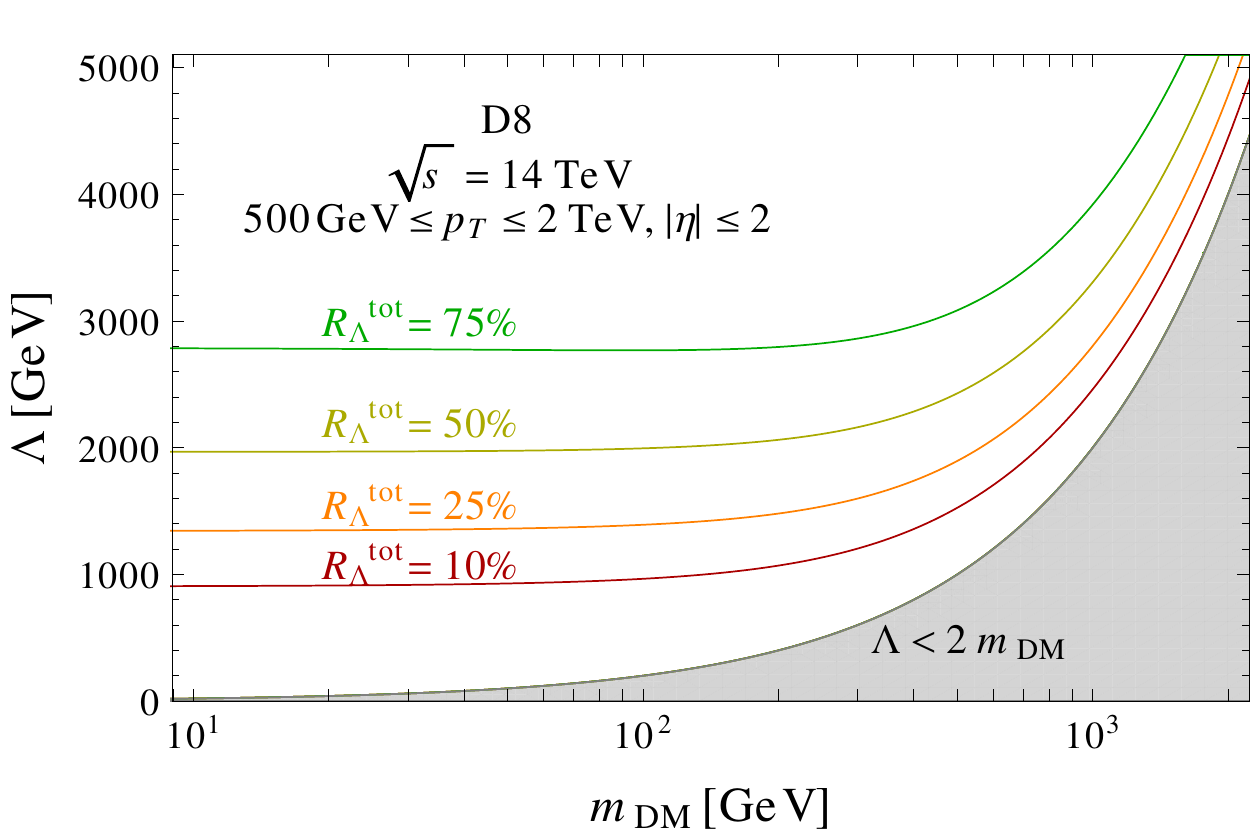}\\
\includegraphics[width=0.45\textwidth]{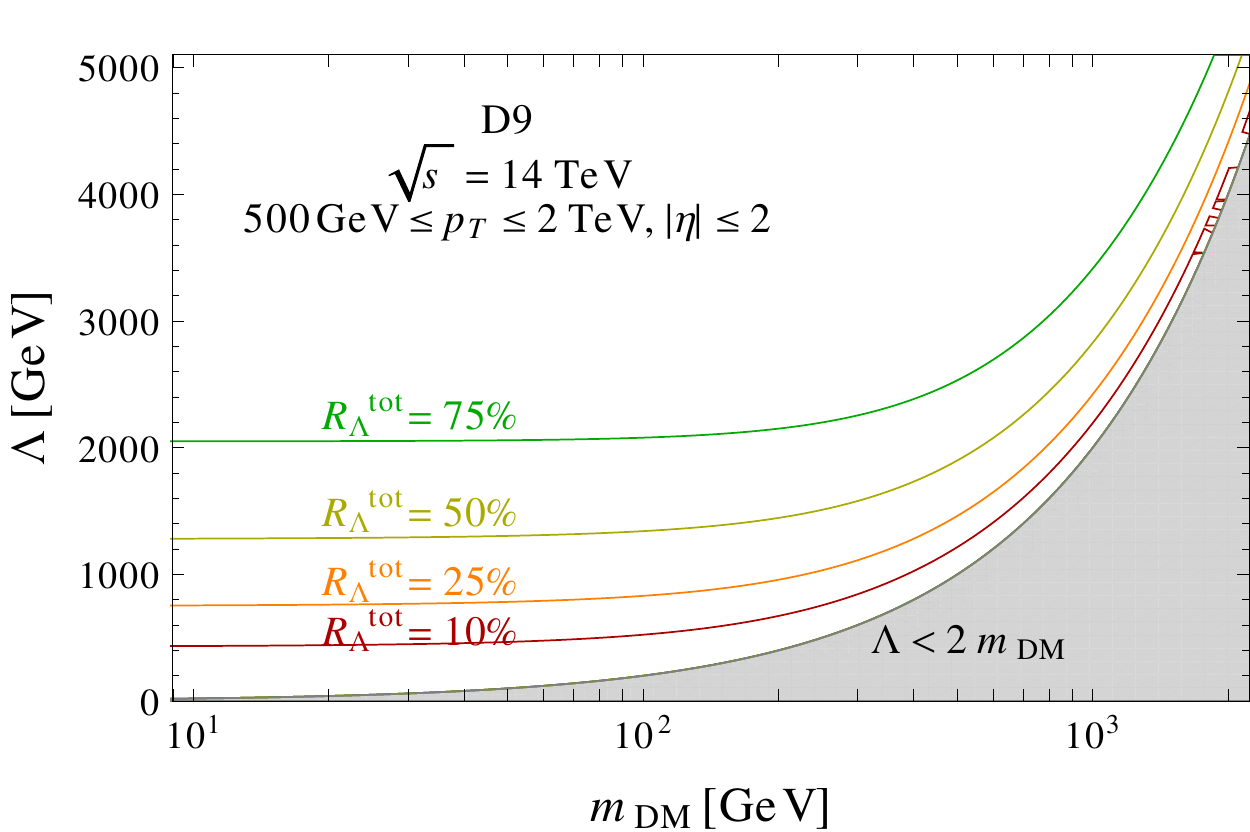}
\hspace{0.5cm}
\includegraphics[width=0.45\textwidth]{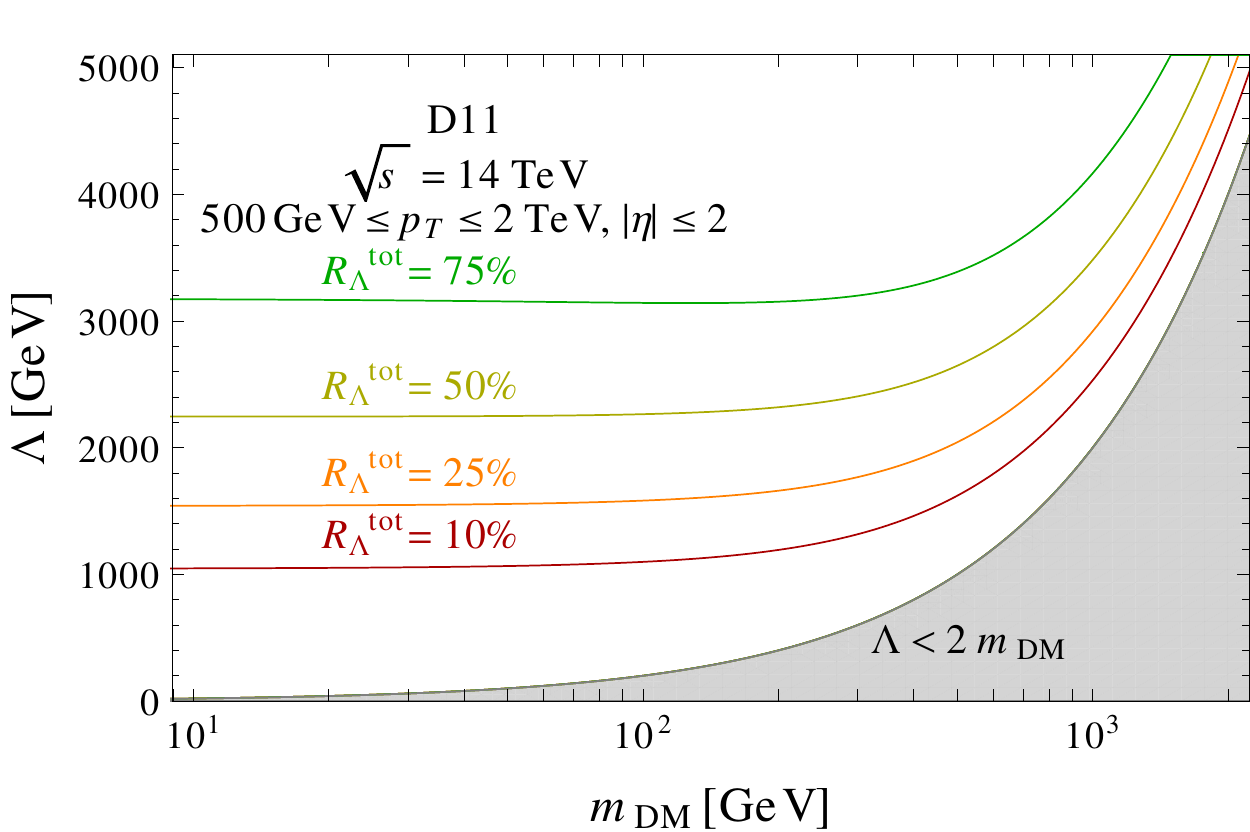}\\
\caption{ \em
Contours for the ratio $R_{\Lambda}^{\rm tot}$, defined in Eq.~(\ref{ratiolambdatot}), on the plane
$(m_{\rm DM}, \Lambda)$, for the different operators. We set $\sqrt{s}=14 {\rm TeV},
|\eta|\leq 2$ and $500 \GeV<p_{\rm T}<2 \TeV$. 
}
\label{fig:RLambdacontours2}
\end{figure}

Next, we turn to study the contours of constant values of the quantity $R_{\Lambda}^{\rm tot}$,
in the plane $(m_{\rm DM}, \Lambda)$. These contour curves for the different operators
are shown in Fig.~\ref{fig:RLambdacontours} for $\sqrt{s}=8$ {\rm TeV} 
and in Fig.~\ref{fig:RLambdacontours2} for $\sqrt{s}=14$ {\rm TeV}.
The requirement that at least 50\% of the events occur with momentum transfer below the cutoff
scale $\Lambda$ requires such a cutoff scale to be above
$\sim 1 {\rm TeV}$ for $\sqrt{s}=8\, {\rm TeV}$, 
or above $\sim 2 {\rm TeV}$ for $\sqrt{s}=14\, {\rm TeV}$. 
Note also that the contours for $D1$--$D4$ differ by the corresponding contours
for $D1'$--$D4'$ by ${\cal O}(1)$ factors, due to the different weighting of the quarks' PDFs.
On the other hand, the experimental bounds
on the scale of the operators $D1$--$D4$ are much lower (of the order of tens of GeV),
as such operators experience an additional suppression of $m_q/\Lambda$.
This means that the bounds on $D1$--$D4$ are not  reliable from the point
of view of EFT validity.

\begin{figure}[t!]
\centering
\includegraphics[width=0.45\textwidth]{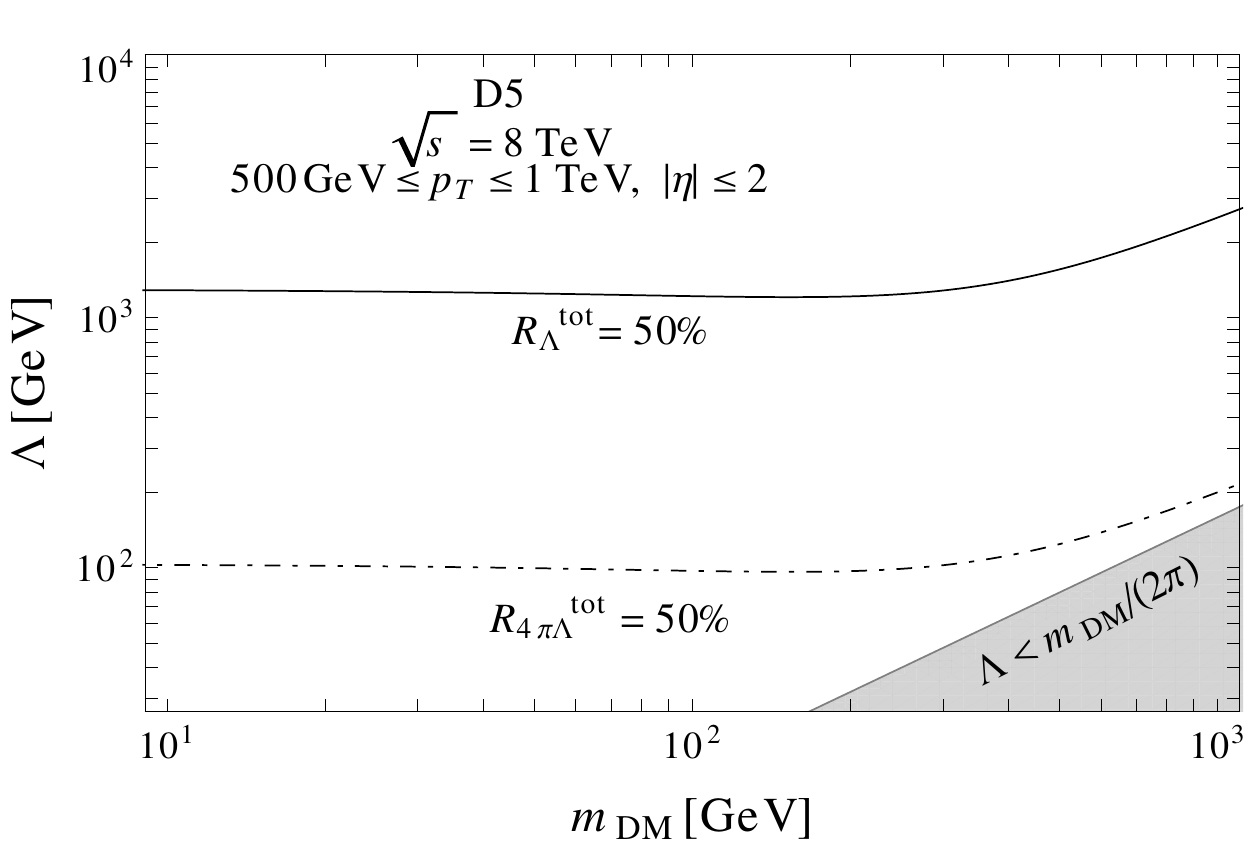}
\hspace{0.5cm}
\includegraphics[width=0.45\textwidth]{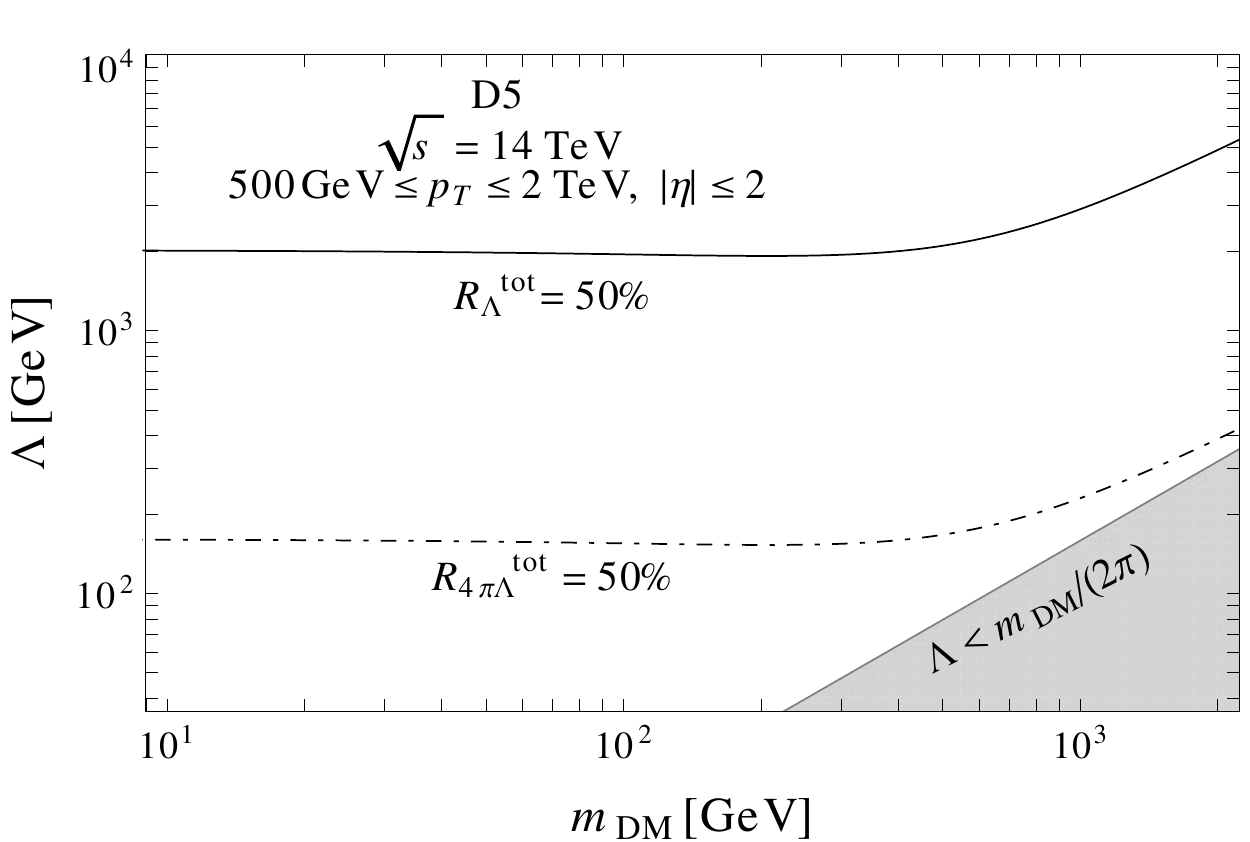}
\caption{ \em
50\% contours for the ratio $R_{\Lambda}^{\rm tot}$ for the operator $D5$, 
varying the cutoff $Q_{\rm tr}<\Lambda$ (solid line) and $Q_{\rm tr}<4\pi\Lambda$ (dot-dashed
line).
We have also shown the region corresponding to $\Lambda<m_{\rm DM}/(2\pi)$ (gray shaded area),
often used as a benchmark for the validity of the EFT.
 We set $\sqrt{s}=8$ {\rm TeV} (left panel) and $\sqrt{s}=14$ {\rm TeV} (right panel).
}
\label{fig:RLambdacontours3}
\end{figure}

We stress once again that the precise definition of a cutoff scale for an EFT is only
possible when the details of the UV completion are known.
The most conservative regime is when the couplings of the UV theory reach their maximal values
allowed by perturbativity. In such a situation, the requirement on the momentum transfer becomes
 $Q_{\rm tr}<4\pi\Lambda$.
We show the effect  of varying the cutoff scale 
 in Fig.~\ref{fig:RLambdacontours3},
for the representative contour $R_\Lambda^{\rm tot}=50\%$ of D5.
As it should be clear, the variation of the cutoff scale is equivalent to a change of the unknown
couplings of the UV theory.
All the operators have very similar results, as the contours scale linearly with the cutoff.
 As a comparison, we show as a shaded area the region
 $\Lambda>m_{\rm DM}/(2\pi)$ often used as a benchmark for the validity of the EFT
 (see Eq.~(\ref{mover2pi})).
The 50\% contour is above such a region, meaning that the parameter space regions of validity of the effective operator approach is smaller than commonly considered.

To close this section let us comment on another 
 question one may ask: 
what is the difference between interpreting data with an effective operator and with
its simplest UV completion?
This question has  already been addressed in  Ref.~\cite{Busoni:2013lha} for the operator $D1'$, by
 studying the ratio of the cross sections obtained with the UV theory and with the effective operator.
 For each of the  operators in Table \ref{table:operators}  one can write a simple UV-complete Lagrangian, see e.g.~Eqs.~(\ref{D1UV})-(\ref{D5UV}).
The very same analysis can be repeated for all the other operators and we checked that
the same qualitative conclusions can be drawn.
In particular, if $\Lambda$ is not larger than a few {\rm TeV}, interpreting the experimental data in
terms of EFT or in terms of a simplified model with a mediator can make a significant difference.

%%%%%%%%%%%%%%%%%%%%%%%%%%%%%%%%%%%%%%%%%%
\section{Comparison with MonteCarlo Simulations}
\label{sec:numerical}
%%%%%%%%%%%%%%%%%%%%%%%%%%%%%%%%%%%%%%%%%%

In order to perform an alternative check of  our analytical results and  to be able to compare to the experimental limits as close as possible, we present 
in this section the results of numerical event simulations. 

\subsection{Simulation and analysis  description}
We made use of  \textsc{MadGraph 5}\cite{mg5}  to simulate $pp$ collisions at $\sqrt{s}=8$ {\rm TeV} and $\sqrt{s}=14$ {\rm TeV}. 
Both PDF sets CTEQ6L1 and MSTW2008LO (discussed in Ref.~\cite{pdf1}) are employed. The PDF choice affects the cross section, but only minimally the acceptance. Hence, the change in contours of $R_\Lambda^{\rm tot}$ is negligible.
Since MSTW2008LO is used for the analytical calculations, this set is also used where direct comparisons between simulation and calculation are shown. For the comparison to the experimental results, CTEQ6L1 is used instead.
Only $u,d,c,s$ quarks were considered, both in the initial and in the final state.

According to the event kinematics we have  evaluated whether or not the conditions of validity discussed in Section \ref{sec:validity} are fulfilled. Specifically, we have  checked if Eqs.~(\ref{cond1}) and 
(\ref{mover2pi}) are fulfilled, that is, if the following condition is satisfied
\be
\Lambda > \frac{Q_{\rm tr}}{\sqrt{g_q g_\chi}} > 2 \frac{m_{\rm DM}}{\sqrt{g_q g_\chi}}\, .
\label{kinconstraint}
\ee
Samples of 20000 events were simulated for each operator, scanning DM mass values of 10, 50, 80, 100, 400, 600, 800 and 1000 GeV and cutoff scales of 250, 500, 1000, 1500, 2000, 2500 and 3000 GeV in the case of $\sqrt{s}=8$ {\rm TeV} collisions. When increasing the collision energy to $\sqrt{s}=14$ {\rm TeV}, the DM mass of 2000 GeV and cutoff scales of 4000 and 5000 GeV were added.

From the simulated samples the fraction of events fulfilling $\Lambda > Q_{\rm tr}/\sqrt{g_q g_\chi}$ for each pair of DM mass and cutoff scale can be evaluated, if one assumes a certain value for the couplings $\sqrt{g_\chi g_q}$ connecting the cutoff scale $\Lambda$ and the mediator mass $M$ via $\Lambda = M/\sqrt{g_q g_\chi}$. As above, $g_q g_\chi$ was assumed to be 1.

\begin{figure}[t!]
\centering
\includegraphics[width=0.55\textwidth]{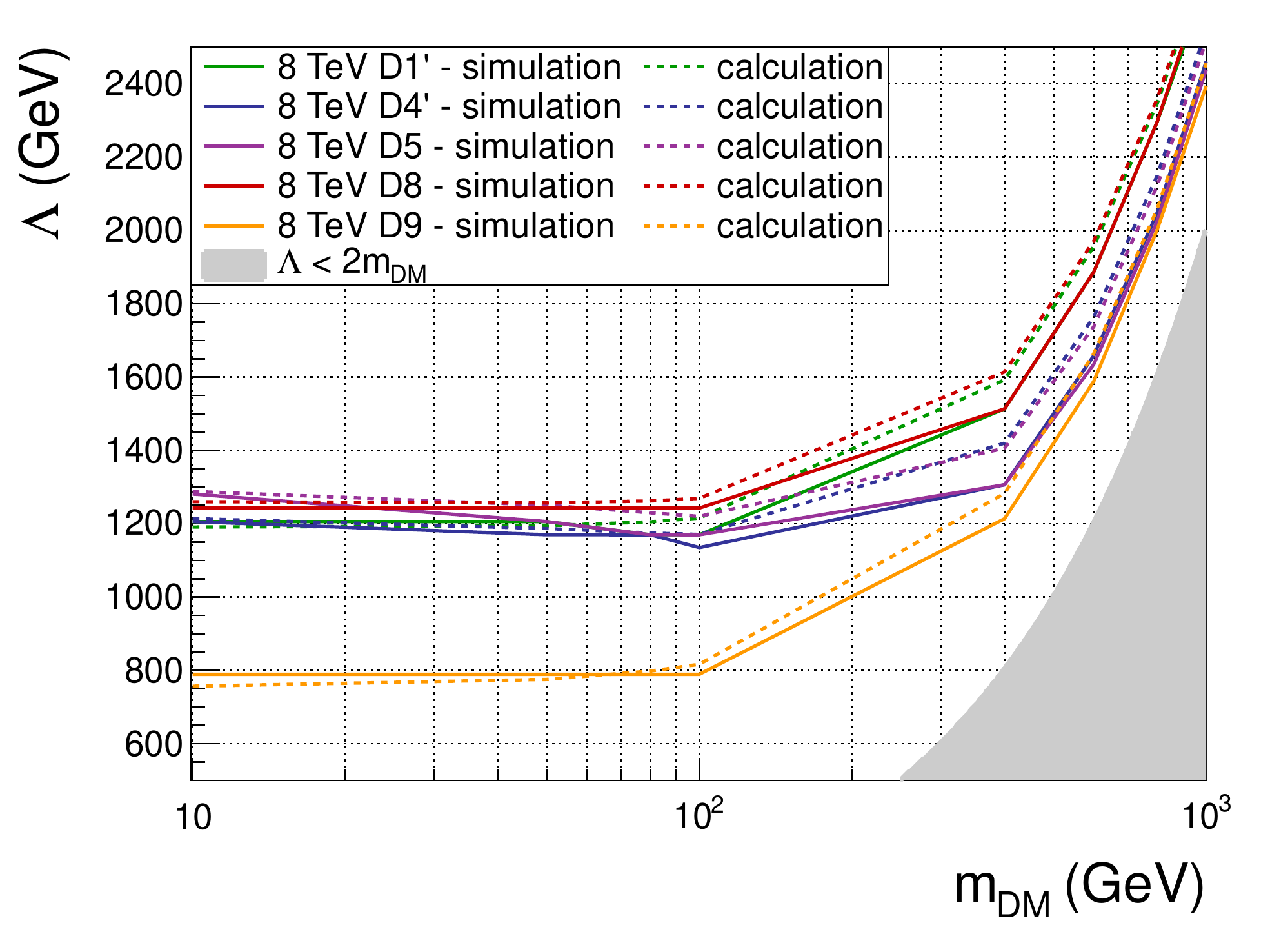}
\caption{ \em
Comparison of the contour $R_\Lambda^{\rm tot} = 50 \%$ for the analytical calculation (dashed line) and the simulation (solid line) for the different operators $D1'$, $D4'$, $D5$, $D8$ and $D9$. The results agree within less than 7 \%.
}
\label{fig:calcsimcomp}
\end{figure}

\subsection{Results}

In order to confirm that analytical and numerical results are in agreement, Figure~\ref{fig:calcsimcomp}  shows a comparison for the operators $D1'$, $D4'$, $D5$, $D8$ and $D9$. The results were obtained for the scenario of one radiated gluon jet above 500 GeV within $|\eta|<2$. The contours of $R_\Lambda^{\rm tot} = 50 \%$ from analytical and numerical evaluation agree within less than 7 \%. The remaining differences could be due to the upper jet $p_{\rm T}$ cut not imposed during event simulation but needed for the analytical calculation, and the details of the fitting procedures.

\begin{figure}[t!]
\centering
\includegraphics[width=0.55\textwidth]{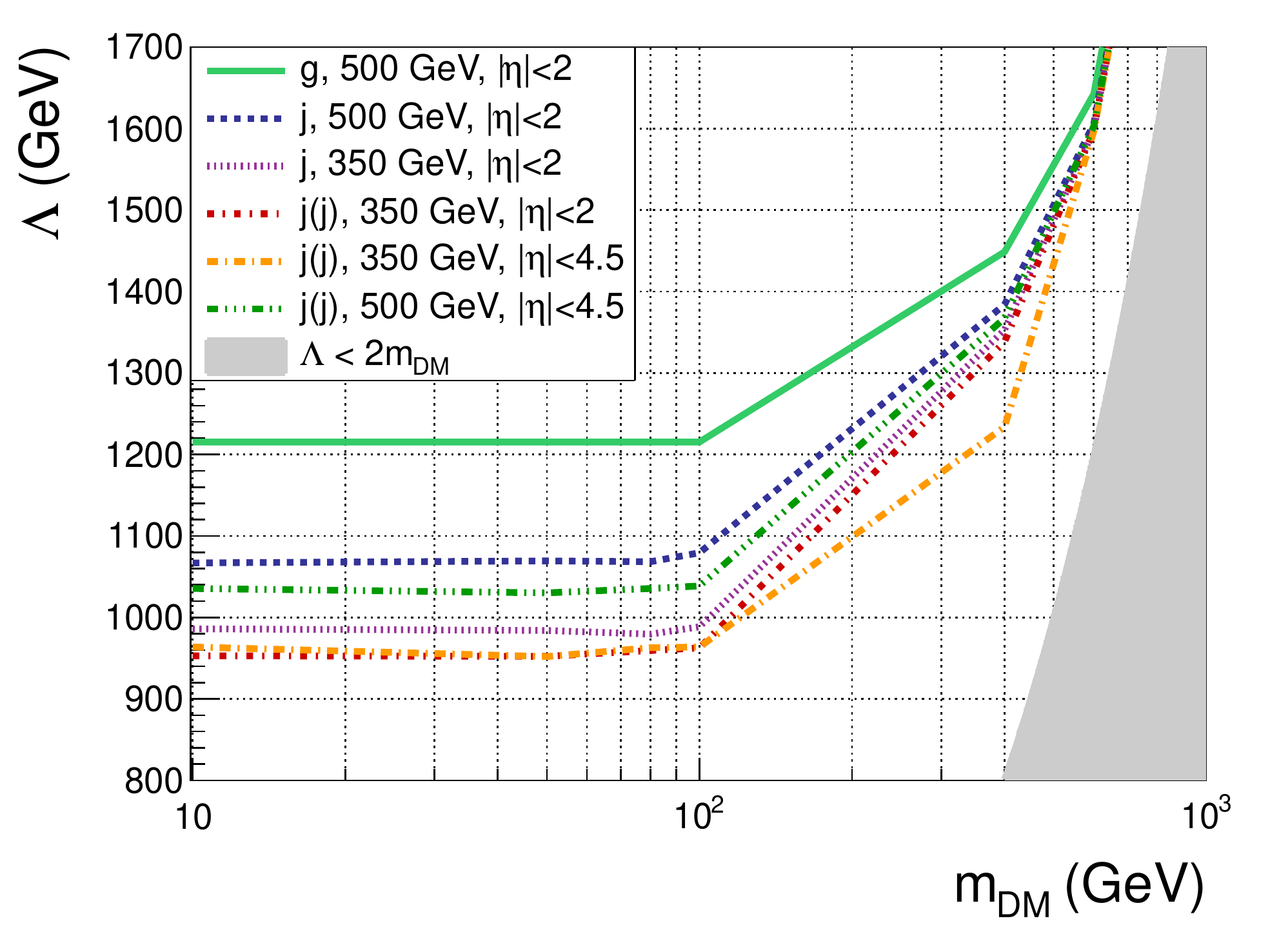}
\caption{ \em
 The changes of the contour of $R_\Lambda^{\rm tot} = 50 \%$ are shown for several variations from the analytically calculated scenario to a scenario close to the cuts used in the ATLAS monojet analysis exemplarily for the operator $D5$ at $\sqrt{s}=8$ TeV. In the legend, ``g'' means only gluon radiation, ``j'' stands for
either quark- or gluon-initiated jets, ``j(j)'' means a second jet is allowed.
}
\label{fig:var_comp}
\end{figure}

Next, we vary the kinematical constraints step by step from the scenario considered in the analytical calculations, namely one radiated gluon jet above 500 GeV within $|\eta|<2$, to a scenario closest to the analysis cuts applied in the ATLAS monojet analysis~\cite{monojetATLAS2}. 
More specifically, the leading jet is allowed to come from either a gluon or a  quark being radiated, the leading jet $p_{\rm T}$ cut is changed from 500 GeV to 350 GeV, a second jet is allowed and its range in $\eta$ is enlarged to $|\eta|< 4.5$. No further cuts are applied at simulation level. 

The effect of the variation of the cuts can be seen in Figure~\ref{fig:var_comp}. Allowing not only for a gluon jet but also taking into account the possibility of a quark jet changes the $R_\Lambda^{\rm tot}$ contours appreciably. The change from lowering the $p_{\rm T}$ of the leading jet has a smaller effect. Allowing for a second jet and enhancing its rapidity range barely changes the $R_\Lambda^{\rm tot}$ contour, especially at large $m_{\rm DM}$ values.

If the collision energy is augmented to $\sqrt{s}=14$ {\rm TeV}, all the $R_{\Lambda}^{\rm tot}$ contours increase. 
As seen for $\sqrt{s}=8$ {\rm TeV}, moving to the scenario closer to the experimental analysis leads to contours that are at most  $\sim30\%$ lower in $\Lambda$.

After having extracted $R_\Lambda^{\rm tot}$ for each WIMP and mediator mass, a curve can be fitted through the points obtained in the plane of $R_\Lambda^{\rm tot}$ and $\Lambda$. The following functional form is used for this purpose

\be
R_\Lambda^{\rm tot}= \left[1-e^{-a\left(\frac{\Lambda-2m_{\rm DM}}{b}\right)^{c}}
\right]\left[1-e^{-d\left(\frac{\Lambda+2m_{\rm DM}}{b}\right)^{e}}
\right]\, .
\label{fit_sim}
\ee
 Further, the parameters are fitted for each DM mass separately. From these fits, the points denoting a cutoff scale where $R_\Lambda^{\rm tot}$ equals  e.g. 50\% can be extracted for each DM mass, and
 the lines of constant $R_\Lambda^{\rm tot}$ can be plotted in the usual limit-setting plane $\Lambda$ vs.~$m_{\rm DM}$.
Table \ref{table:fit_sim} collects the values of the fitting parameters
for all operators except D12-D14, for which no experimental analysis exists.

\begin{table}[t]
\centering
\begin{tabular}{cc}
$\sqrt{s}=8$ {\rm TeV} & $\sqrt{s}=14$ {\rm TeV}\\ 
\begin{tabular}{| c || c | c | c | c | c |}
  \hline                        
  Operator & a & b & c & d & e \\ \hline
  D1  & 1.32 & 787.13 & 1.39 & 1.08 & 1.53 \\ \hline
  D1' & 1.30 & 1008.25& 1.49 & 0.77 & 1.83 \\ \hline
  D4  & 1.65 & 702.93 & 1.14 & 0.65 & 1.75 \\ \hline
  D4' & 1.51 & 859.83 & 1.22 & 0.48 & 1.92 \\ \hline
  D5  & 1.54 & 816.83 & 1.18 & 0.50 & 1.85 \\ \hline
  D8  & 1.23 & 964.62 & 1.50 & 0.91 & 1.59 \\ \hline
  D9  & 1.43 & 681.92 & 1.15 & 1.02 & 1.35 \\ \hline
  D11 & 1.23 & 1002.33& 1.49 & 0.82 & 1.69 \\ \hline
\end{tabular}
\hspace{2cm}&
\begin{tabular}{| c || c | c | c | c | c |}
  \hline                        
  Operator & a & b & c & d & e \\ \hline
  D1  & 0.89 & 1017.37& 1.45 & 1.28 & 1.24 \\ \hline
  D1' & 0.43 & 909.66 & 1.59 & 0.53 & 1.37 \\ \hline
  D4  & 1.23 & 996.82 & 1.25 & 0.80 & 1.48 \\ \hline
  D4' & 0.76 & 982.75 & 1.33 & 0.37 & 1.63 \\ \hline
  D5  & 0.78 & 894.86 & 1.25 & 0.39 & 1.54 \\ \hline
  D8  & 0.48 & 945.09 & 1.55 & 0.74 & 1.24 \\ \hline
  D9  & 0.91 & 891.65 & 1.21 & 1.23 & 1.04 \\ \hline
  D11 & 0.68 & 1250.49& 1.58 & 0.81 & 1.35 \\ \hline
\end{tabular}
\end{tabular}
\caption{\em Coefficient for the fitting functions for $R_\Lambda^{\rm tot}$ in Eq.~(\ref{fit_sim}), in the cases $\sqrt{s}=$ 8 and 
14 {\rm TeV}.
The fitting functions describe processes where quarks and/or gluons are radiated, 
the final state contains 1 or 2 jets, where the leading jet has minimum $p_{\rm T}$ of 350 GeV while the second jet is allowed to be within $|\eta|< 4.5$. See text for
further details.
}
\label{table:fit_sim}
\end{table}

%%%%%%%%%%%%%%%%%%%%%%%%%%%%%%%%%%%%%%%%%%

\section{Implications of the limited validity of EFT in DM searches at LHC}
\label{sec:interp}

\begin{figure}[t!]
\centering
\includegraphics[width=0.45\textwidth]{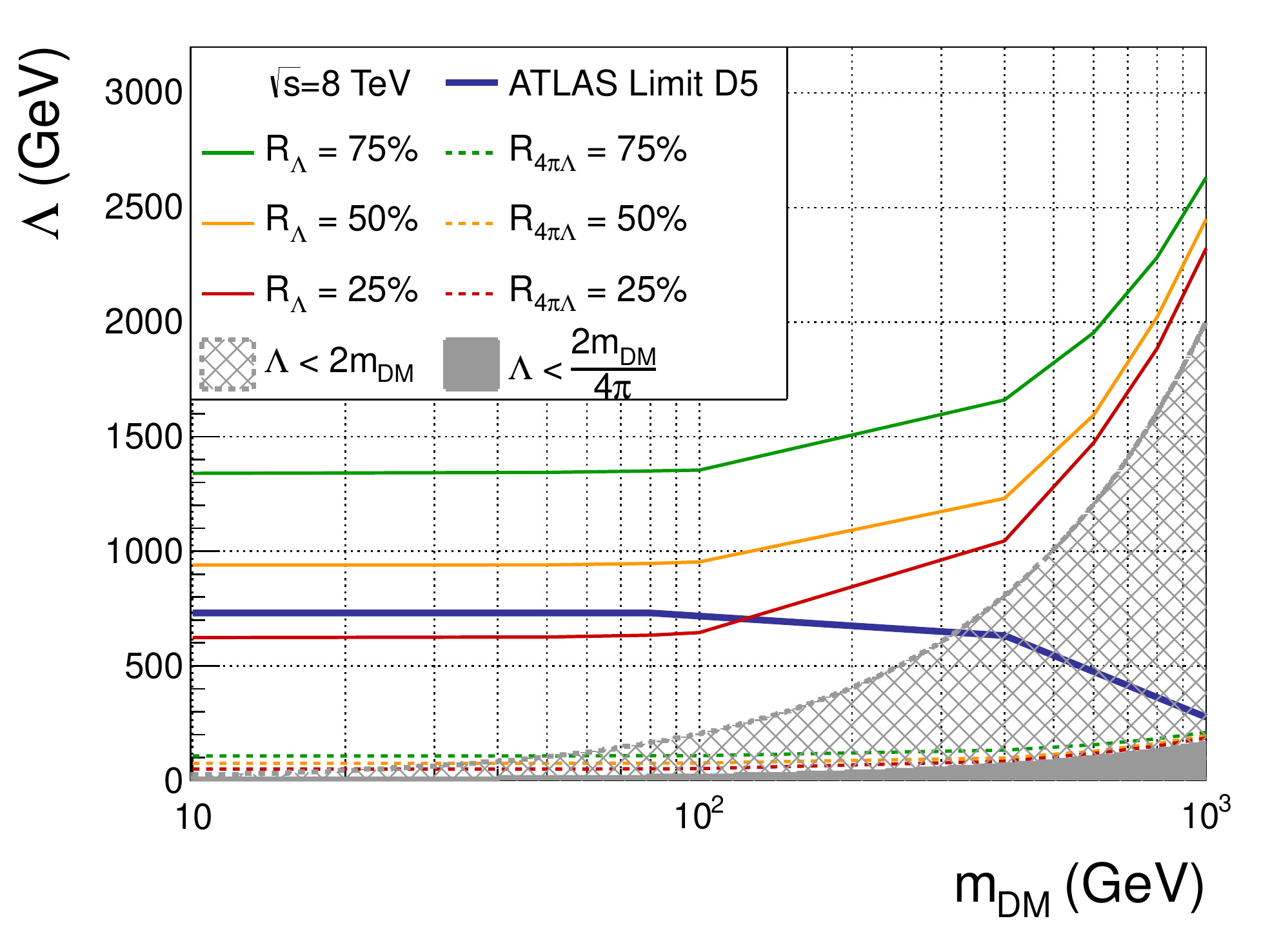}
\hspace{0.5cm}
\includegraphics[width=0.45\textwidth]{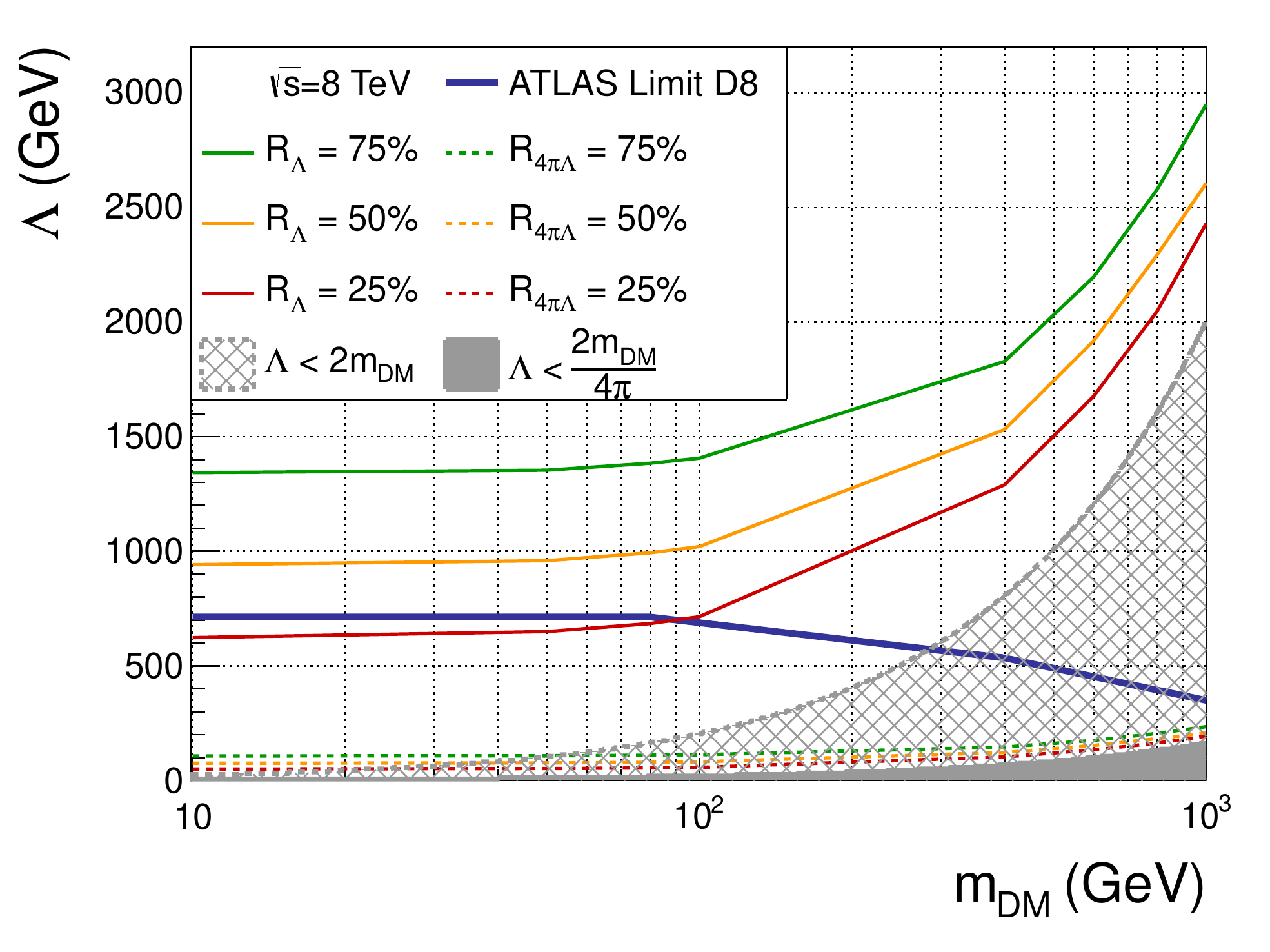}\\
\includegraphics[width=0.45\textwidth]{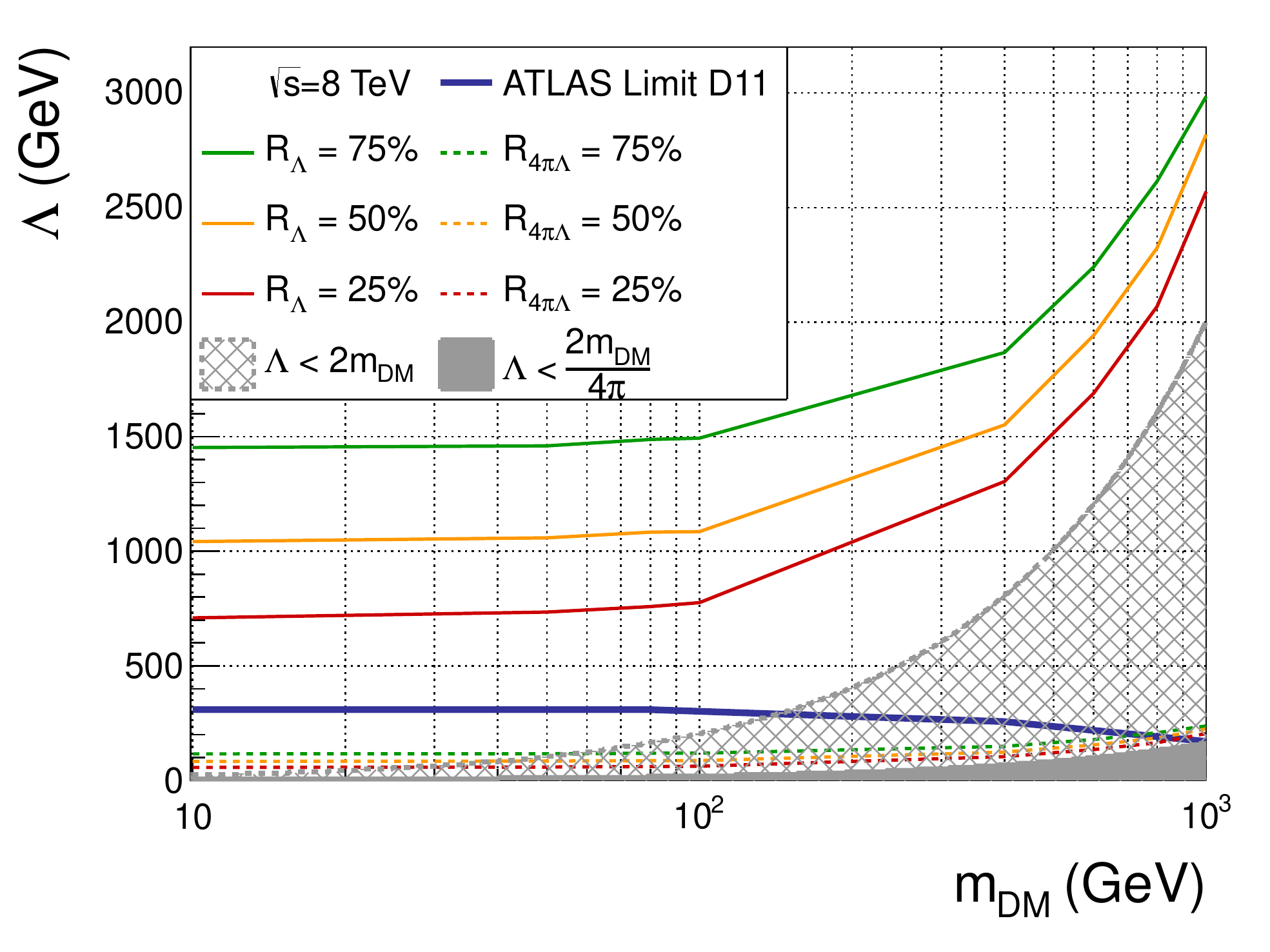}
\caption{\em
25\%, 50\% and 75\% contours for the ratio $R_{\Lambda}^{\rm tot}$, compared to the experimental limits from ATLAS~\cite{monojetATLAS2} (blue line). Also indicated are the contours of $R_{\Lambda}^{\rm tot}$ in the extreme case when setting the couplings $\sqrt{g_q g_\chi}=4\pi$ (dashed lines). Results are shown for different operators: D5 (upper left panel), D8 (upper right panel) and D11 (lower panel).
 }
\label{fig:expcomp}
\end{figure}

Figure \ref{fig:expcomp} shows the experimental limits obtained from the ATLAS monojet  analysis
 \cite{monojetATLAS2} in the plane $(\Lambda$, $m_{\rm DM}$), for the opearators D5, D8 and D11.
The contours of $R_{\Lambda}^{\rm tot}$ for 25\%, 50\% and 75\% are superimposed. The experimental limits are placed in a region where about 30\% of the events can be expected to fulfill the EFT conditions - the exact number depends on the operator considered. Especially the limit on the gluon operator $D11$ seems questionnable. For comparison, dashed lines show the contours of $R_{\Lambda}^{\rm tot}$ for the extreme case of  couplings $\sqrt{g_q g_\chi}=4\pi$, presenting the limiting case for which the theory is still considered perturbative.

Unfortuntately, there is no possibility to measure $Q_{\rm tr}$ in data, on an event-by-event basis. 
So the information on what is the fraction of the events to cut out
comes from  analytical computations or a numerical simulation, as we explained in this paper.
To assess the impact of the limited validity of the EFT on the current collider bounds, we adopt the  procedure that  relies on the assumption that the $p_{\rm T}$ (or MET) distributions with the $Q_{\rm tr}$ cut are simply a rescaling of those without the cut. 
A more refined study should account for possible kinematic shape changes with the jet transverse momentum and/or missing energy and DM mass\footnote{Preliminary studies
 indicate that the method adopted in this paper is quite reasonable for   cuts with $Q_{\rm tr}<750$ GeV or weaker \cite{private}.}.

\begin{figure}[t!]
\centering
\includegraphics[width=0.45\textwidth]{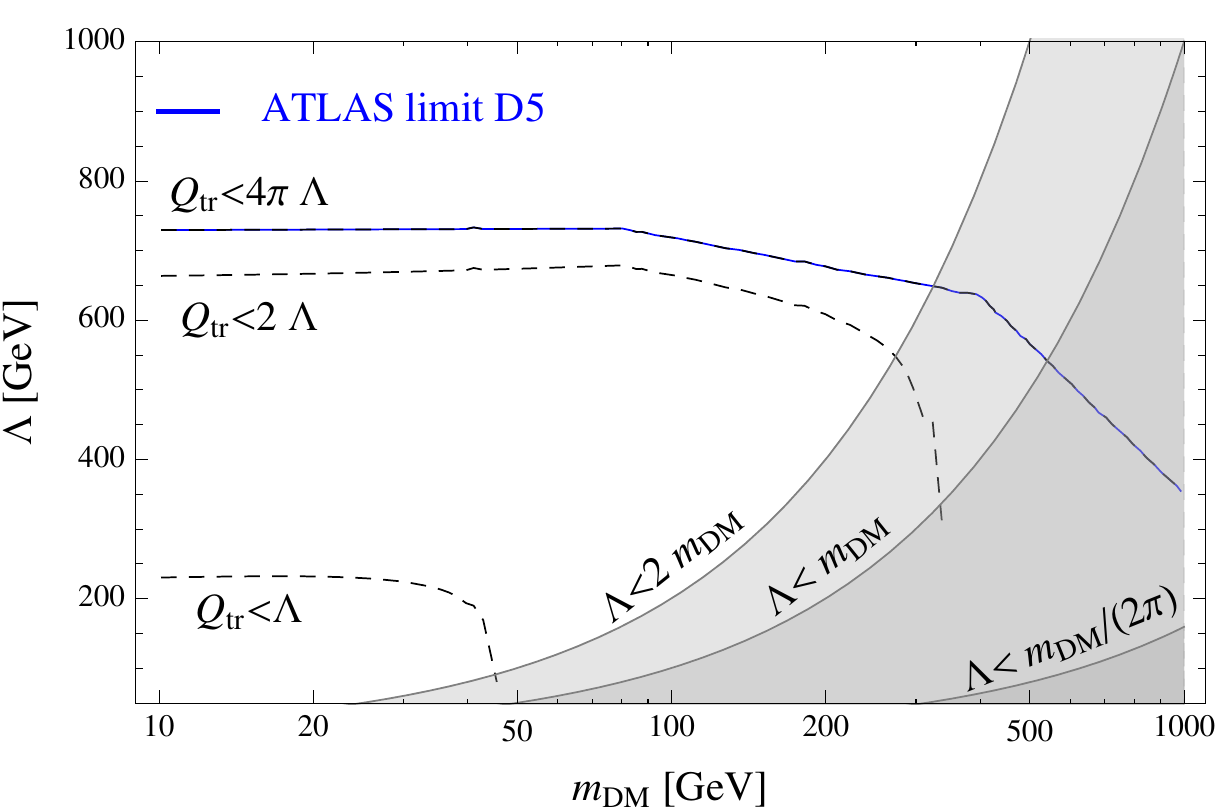}
\hspace{0.5cm}
\includegraphics[width=0.45\textwidth]{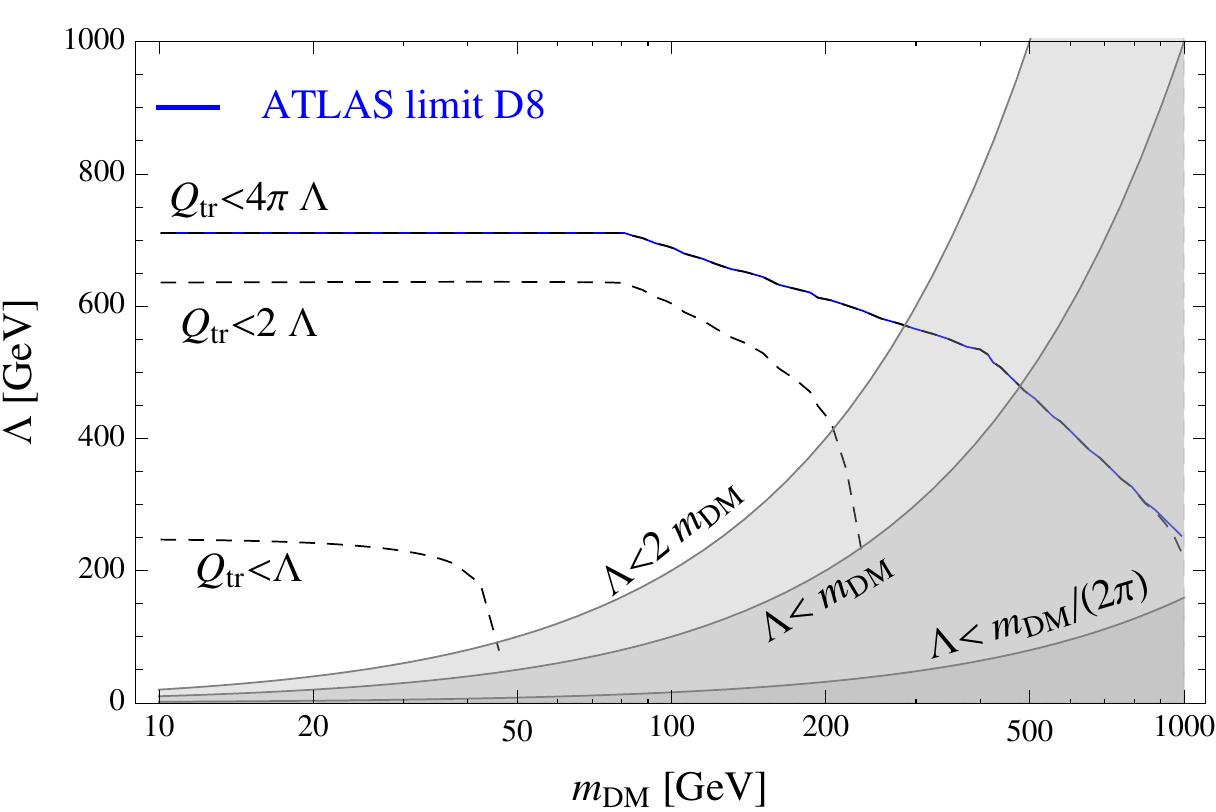}\\
\includegraphics[width=0.45\textwidth]{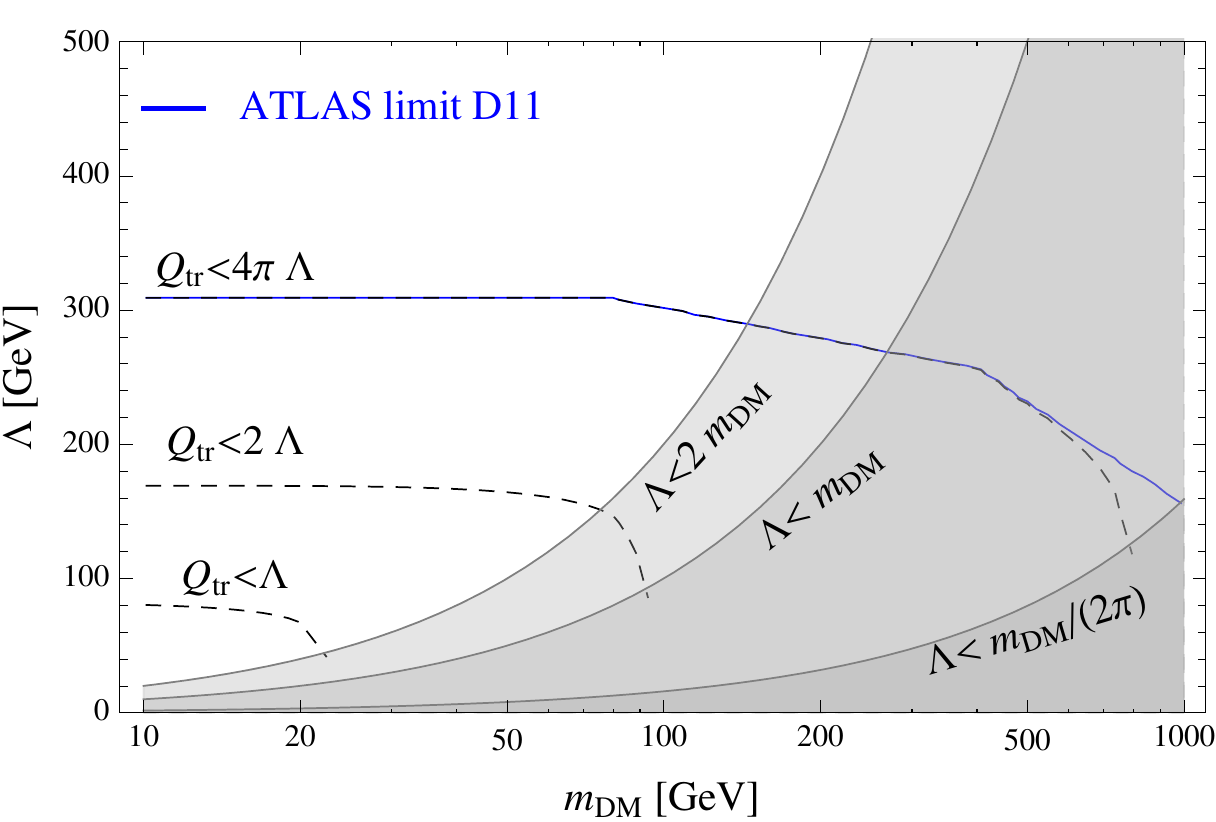}
\caption{ \em
The experimental limits by ATLAS \cite{monojetATLAS2} on the suppression scale $\Lambda$
are shown as solid blue lines.
The updated limits  taking into account EFT validity are shown as dashed black lines,
for $Q_{\rm tr}<\Lambda, 2\Lambda, 4\pi\Lambda$,
corresponding to different choices of the UV couplings: $\sqrt{g_qg_\chi}=1, 2, 4\pi$, respectively.
The corresponding kinematical constraints (Eq.~(\ref{kinconstraint})) are denoted by gray bands.
The different plots refer to different operators: D5  (upper left panel), D8 (upper right panel)
and D11 (lower panel). 
}
\label{fig:newlimit}
\end{figure}

Very naively, neglecting the statistical and systematical uncertainties, the number of signal events in a given EFT model has to be
less than the experimental observation, $N_{\rm signal}(\Lambda, m_{\rm DM})< N_{\rm expt}$.
The cross section due to an operator of mass dimension $d$  scale like $\Lambda^{-2(d-4)}$, so
 $N_{\rm signal} (\Lambda , m_{\rm DM})=\Lambda^{-2(d-4)} \tilde{N}_{\rm signal}(m_{\rm DM})$,
and the experimental lower bound in the scale of the operator becomes
\be
\Lambda> \left[\tilde N_{\rm signal}(m_{\rm DM})/N_{\rm exp}\right]^{1/[2(d-4)] }\equiv \Lambda_{\rm expt.}\,.
\ee
Now, if we do not consider any information about the shapes of the $p_{\rm T}$ or MET distributions,
the experimental bound only comes from the total number of events passing given cuts.
The fact that a fraction of the events involve a transfer momentum exceeding the cutoff scale of the EFT
means that the number of signal events  for placing a limit gets reduced by a factor  $R_\Lambda^{\rm tot}$.
Therefore,  actually $N_{\rm signal}(\Lambda, m_{\rm DM}) \to R_{\Lambda}^{\rm tot}(m_{\rm DM}) N_{\rm signal}(\Lambda, m_{\rm DM})$, 
so the new limit is found by solving the implicit equation
\be
\Lambda> [ R_{\Lambda}^{\rm tot}( m_{\rm DM})]^{1/2[(d-4)]} [N_{\rm signal}(m_{\rm DM})/N_{\rm exp}]^{1/[2(d-4)]}= [R_{\Lambda}^{\rm tot}( m_{\rm DM})]^{1/[2(d-4)]}  \Lambda_{\rm expt}
\label{eq:implnewlim}
\ee
and it turns out to be  weaker than $\Lambda_{\rm expt}$.
In Fig.~\ref{fig:newlimit} we show the new limits for the dim-6 operators D5, D8 and the dim-7 operator D11, for the conditions $Q_{\rm tr}<\Lambda, 2\Lambda, 4\pi\Lambda$,
corresponding different choices of the UV couplings: $\sqrt{g_qg_\chi}=1, 2, 4\pi$, respectively.
The curves are obtained solving Eq. \ref{eq:implnewlim} with $R_{\Lambda}^{\rm tot},R_{2\Lambda}^{\rm tot},R_{4\pi\Lambda}^{\rm tot}$ respectively.
The  ATLAS bound reported is the 90\%CL observed limit.
The functions $R_\Lambda^{\rm tot}$ used are taken from the fitting functions described in Table \ref{table:fit_sim}, which include both quark and gluon jets, and the same cuts as the ``Signal Region 3''
used by ATLAS.
As expected, the weaker is the condition on $Q_{\rm tr}$, the more the new limits approach the ATLAS bound.
In the case of extreme couplings $\sqrt{g_qg_\chi}=4\pi$, the condition on the momentum transfer is 
very conservative $Q_{\rm tr}<4\pi\Lambda$. For D5 and D8, the new limit is indisinguishable
from the ATLAS one, meaning that the experimental results are safe from the EFT point of view,
in this limiting situation. For D11, even for extreme values of the couplings, the bound at large DM masses must be corrected.
In general, for couplings of order one, the limits which are safe from the EFT point of view are appreciably
weaker than those reported. We encourage the experimental collaborations to take this point into account when publishing their limits.

%%%%%%%%%%%%%%%%%%%%%%%%%%%%%%%%%%%%%%%%%%
\section{Conclusions}
\label{sec:conclusions}
%%%%%%%%%%%%%%%%%%%%%%%%%%%%%%%%%%%%%%%%%%
The search for DM is one of the main targets of LHC analyses. In this paper we have continued our previous investigation to assess the validity of the EFT commonly used in interpreting such searches. Following Ref. \cite{Busoni:2013lha}, we have studied the quantity $R_\Lambda^{\rm tot}$
(see Eq.(~\ref{ratiolambdatot}), which quantifies the error made when using effective operators to describe processes with very high momentum transfer. Our criterion indicates up to what cutoff energy scale the effective description is valid, depending on the DM mass and couplings.
We have performed the analysis for the full list of EFT operators,  connecting fermion DM particles and quarks or gluons,  used by the ATLAS and CMS collaborations and originated from the exchange of heavy mediators in the $s$-channel. 
We have also extended our analysis to the case of $\sqrt{s}=$ 14 {\rm TeV}. Furthermore,
we have validated our analytical results by performing  numerical event simulations which reproduce the experimental situation 
in the  closest possible  way. Our results indicate that the range of validity of the EFT is significantly limited in the parameter space
$(\Lambda,m_{\rm DM})$. While our findings are valid for the $s$-channel, a similar analysis is under way for the $t$-channel \cite{thomas} where similar results are obtained.

Does it mean that the EFT is not the best tool to interpret the current LHC data of DM searches? The answer is yes and no. On the negative  side, our results clearly cry out for an overcoming of the EFT, most possibly through 
identifying  a handful of classes of models (able to reproduce the EFT operators in the heavy mediator limit); this would allow a consistent
analysis of the current and future LHC data by consistently taking into account the role played by the mediator. On the positive side, keep working with the EFT allows to avoid the overwhelming model-dependence generated by the many DM models proposed so far. 
Nonetheless,  as we have shown in section \ref{sec:interp}, the price to pay 
is  a deterioration of the limits presented so far.

\section*{Acknowledgments}
We thank A. Brennan, C. Doglioni, G. Iacobucci, T. Jacques, S. Schramm and S. Vallecorsa for many interesting conversations.
ADS acknowledges partial support from the  European Union FP7  ITN INVISIBLES 
(Marie Curie Actions, PITN-GA-2011-289442).
JG acknowledges partial support from UNIGE and SNF grant 200020-144493, ``High-Energy Hadron Interactions: ATLAS at the CERN LHC''.

\appendix
%%%%%%%%%%%%%%%%%%%%%%%%%%%%%%%%%%%%%%%%%%
\section{Three-body Cross Sections}
\label{app:crosssect}
%%%%%%%%%%%%%%%%%%%%%%%%%%%%%%%%%%%%%%%%%%

\subsection{Generalities}

In this Appendix we show the details of the calculations of the tree-level cross sections for
 the hard scattering process $f(p_1)+\bar f(p_2)\to \chi(p_3)+\chi(p_4)+g(k)$, where 
 $f$ is either a quark (operators D1-D10) or a gluon (D11-D14), and the final
 gluon is  emitted from the initial state.

The differential
 cross section    is generically given by
\begin{equation}
d\hat\sigma=\frac{\sum\overline{|{\cal M}|^2}}{4(p_1\cdot p_2)}
\de \Phi_3\,,
\end{equation}
where  the three-body phase space is
\be
\de \Phi_3=
(2\pi)^4\delta^{(4)}(p_1+p_2-p_3-p_4-k)\frac{{\rm d}\textbf{p}_3}{(2\pi)^32p_3^0}
\frac{{\rm d} \textbf{p}_4}{(2\pi)^32p_4^0}\frac{{\rm d}\textbf{k}}{(2\pi)^32k^0}\,.
\ee

\subsection{Matrix Elements}

In the limit of massless light quarks, they have definite helicity and it makes no difference
for the cross sections whether there is $q$ or $\gamma^5 q$ in the operator. Therefore
the following identifications between pairs of operators hold:
\be
 D1' \leftrightarrow D3' , \qquad D2' \leftrightarrow D4',\qquad D5 \leftrightarrow D7, \qquad 
 D6 \leftrightarrow D8, \qquad D9 \leftrightarrow D10\,,
\label{eqopcorr}
\ee
while the ``primed'' and ``unprimed'' operators are related as in Eq.~(\ref{primeunprime}).
For definiteness, we choose to work with $D1', D4', D5, D8, D9$ and $D11-D14$.

The amplitudes are given by
\bea
{\cal M}_{D1'}&=&-i g_s \frac{1}{\Lambda^2} \epsilon_{\mu}^{*a}(k) \left[\frac{\bar{v}(p_2)({\cancel p_1}-\cancel{k})\gamma^{\mu}T^a u(p_1)}{(p_1-k)^2}-\frac{\bar{v}(p_2)\gamma^{\mu}T^a({\cancel p_2}-\cancel{k})u(p_1)}{(p_2-k)^2}\right]\bar{u}(p_3)v(p_4)\,,\\
{\cal M}_{D4'}&=&-i g_s \frac{1}{\Lambda^2} \epsilon_{\mu}^{*a}(k) \left[\frac{\bar{v}(p_2)\gamma^{5}({\cancel p_1}-\cancel{k})\gamma^{\mu}T^a u(p_1)}{(p_1-k)^2}-\frac{\bar{v}(p_2)\gamma^{\mu}T^a({\cancel p_2}-\cancel{k})\gamma^{5}u(p_1)}{(p_2-k)^2}\right] \nn\\
&&\times \bar{u}(p_3)\gamma^{5}v(p_4)\,,\\
{\cal M}_{D5}&=&-i g_s \frac{g_{\nu\rho}}{\Lambda^2} \epsilon_{\mu}^{*a}(k) 
\left[\frac{\bar{v}(p_2)\gamma^{\nu}({\cancel p_1}-\cancel{k})\gamma^{\mu}T^a u(p_1)}{(p_1-k)^2}-\frac{\bar{v}(p_2)\gamma^{\mu}T^a({\cancel p_2}-\cancel{k})\gamma^{\nu}u(p_1)}{(p_2-k)^2}\right]
\nn\\
&&\times \bar{u}(p_3)\gamma^{\rho}v(p_4)\,,\\
{\cal M}_{D8}&=&-i g_s \frac{g_{\nu\rho}}{\Lambda^2} \epsilon_{\mu}^{*a}(k)
\left[\frac{\bar{v}(p_2)\gamma^{\nu}\gamma^{5}({\cancel p_1}-\cancel{k})\gamma^{\mu}T^a u(p_1)}{(p_1-k)^2}-\frac{\bar{v}(p_2)\gamma^{\mu}T^a({\cancel p_2}-\cancel{k})\gamma^{\nu}\gamma^{5}u(p_1)}{(p_2-k)^2}\right]\nn\\
&&\times \bar{u}(p_3)\gamma^{\rho}\gamma^{5}v(p_4)\,,\\
{\cal M}_{D9}&=&-i\frac{g_s}{16} \frac{g_{\mu\rho}g_{\nu\sigma}}{\Lambda^2} \epsilon_{\alpha}^{*a}(k)
\left[\frac{\bar{v}(p_2)\sigma^{\mu \nu}({\cancel p_1}-\cancel{k})\gamma^{\alpha}T^a u(p_1)}{(p_1-k)^2}-\frac{\bar{v}(p_2)\gamma^{\alpha}T^a({\cancel p_2}-\cancel{k})\sigma^{\mu \nu}u(p_1)}{(p_2-k)^2}\right]
\nn\\
&&\times\bar{u}(p_3)\sigma^{\rho \sigma}v(p_4)\,,\\
{\cal M}_{D11}&=& \frac{g_s^3}{4\pi} \frac{1}{\Lambda^3} f_{abc} \epsilon_{\mu}(p_1) \epsilon_{\nu}(p_2) \epsilon_{\rho}^{*}(k) \bar{u}(p_3)v(p_4)\nn\\
&&\hspace{-0.5cm} \left[\frac{(g^{\mu\sigma}(2p_1-k)^\rho+g^{\rho\sigma}(2k-p_1)^\mu-g^{\mu\rho}(k+p_1)^\sigma)((p_1-k)^\nu p_{2 \sigma}-(p_1-k)\cdot p_2g_\sigma^\nu)}{(p_1-k)^2}\right. \nn\\
&&\hspace{-0.5cm} -\frac{(g^{\nu\sigma}(2p_2-k)^\rho+g^{\rho\sigma}(2k-p_2)^\nu-g^{\nu\rho}(k+p_2)^\sigma)((p_2-k)^\mu p_{1 \sigma}-(p_2-k)\cdot p_1g_\sigma^\mu)}{(p_2-k)^2} \nn\\
&&\hspace{-0.5cm} -\frac{(g^{\mu\nu}(p_1-p_2)^\sigma+g^{\nu\sigma}(p_1+2p_2)^\mu-g^{\mu\sigma}(2p_1+p_2)^\nu)((p_1+p_2)^\rho k_\sigma-k\cdot(p_1+p_2) g_\sigma^\rho)}{(p_1+p_2)^2} \nn\\
&&\hspace{-0.5cm}\left. +g^{\mu\nu}(p_1-p_2)^\rho+g^{\nu\rho}(p_2+k)^\mu-g^{\mu\rho}(k+p_1)^\nu \right]\,,
\eea
\bea
{\cal M}_{D12}&=& i \frac{g_s^3}{4\pi} \frac{1}{\Lambda^3} f_{abc} \epsilon_{\mu}(p_1) \epsilon_{\nu}(p_2) \epsilon_{\rho}^{*}(k) \bar{u}(p_3)\gamma^5v(p_4)\nn\\
&&\hspace{-0.5cm} \left[\frac{(g^{\mu\sigma}(2p_1-k)^\rho+g^{\rho\sigma}(2k-p_1)^\mu-g^{\mu\rho}(k+p_1)^\sigma)((p_1-k)^\nu p_{2 \sigma}-(p_1-k)\cdot p_2g_\sigma^\nu)}{(p_1-k)^2}\right. \nn\\
&&\hspace{-0.5cm} -\frac{(g^{\nu\sigma}(2p_2-k)^\rho+g^{\rho\sigma}(2k-p_2)^\nu-g^{\nu\rho}(k+p_2)^\sigma)((p_2-k)^\mu p_{1 \sigma}-(p_2-k)\cdot p_1g_\sigma^\mu)}{(p_2-k)^2} \nn\\
&&\hspace{-0.5cm} -\frac{(g^{\mu\nu}(p_1-p_2)^\sigma+g^{\nu\sigma}(p_1+2p_2)^\mu-g^{\mu\sigma}(2p_1+p_2)^\nu)((p_1+p_2)^\rho k_\sigma-k\cdot(p_1+p_2) g_\sigma^\rho)}{(p_1+p_2)^2} \nn\\
&&\hspace{-0.5cm}\left. +g^{\mu\nu}(p_1-p_2)^\rho+g^{\nu\rho}(p_2+k)^\mu-g^{\mu\rho}(k+p_1)^\nu \right]\,,\\
{\cal M}_{D13}&=& - \frac{g_s^3}{4\pi} \frac{1}{\Lambda^3} f_{abc} \epsilon_{\mu}(p_1) \epsilon_{\nu}(p_2) \epsilon_{\rho}^{*}(k) \bar{u}(p_3)v(p_4)\nn\\
&&\hspace{-0.5cm} \left[\frac{(g_\sigma^{\mu}(2p_1-k)^\rho+g_\sigma^{\rho}(2k-p_1)^\mu-g^{\mu\rho}(k+p_1)_\sigma)(\epsilon^{\sigma\nu\eta\chi}p_{2 \eta}(p_1-k)_\chi))}{(p_1-k)^2}\right. \nn\\
&&\hspace{-0.5cm} +\frac{(g_\sigma^{\nu}(2p_2-k)^\rho+g_\sigma^{\rho}(2k-p_2)^\nu-g^{\nu\rho}(k+p_2)_\sigma)(\epsilon^{\sigma\mu\eta\chi}p_{1 \eta}(p_2-k)_\chi)}{(p_2-k)^2} \nn\\
&&\hspace{-0.5cm}
 +\frac{(g^{\mu\nu}(p_1-p_2)_\sigma+g_\sigma^{\nu}(p_1+2p_2)^\mu-g_\sigma^{\mu}(2p_1+p_2)^\nu)(\epsilon^{\rho\eta\sigma\chi}k_{\eta}(p_1+p_2)_\chi)}{(p_1+p_2)^2}\nn\\
&&\left.-\epsilon^{\mu\nu\rho\sigma}(p_1+p_2-k)_\sigma \right] \,,\\
{\cal M}_{D14}&=& -i \frac{g_s^3}{4\pi} \frac{1}{\Lambda^3} f_{abc} \epsilon_{\mu}(p_1) \epsilon_{\nu}(p_2) \epsilon_{\rho}^{*}(k) \bar{u}(p_3)\gamma^5v(p_4)\nn\\
&&\hspace{-0.5cm} \left[\frac{(g_\sigma^{\mu}(2p_1-k)^\rho+g_\sigma^{\rho}(2k-p_1)^\mu-g^{\mu\rho}(k+p_1)_\sigma)(\epsilon^{\sigma\nu\eta\chi}p_{2 \eta}(p_1-k)_\chi))}{(p_1-k)^2}\right. \nn\\
&&\hspace{-0.5cm} +\frac{(g_\sigma^{\nu}(2p_2-k)^\rho+g_\sigma^{\rho}(2k-p_2)^\nu-g^{\nu\rho}(k+p_2)_\sigma)(\epsilon^{\sigma\mu\eta\chi}p_{1 \eta}(p_2-k)_\chi)}{(p_2-k)^2} \nn\\
&&\hspace{-0.5cm}  
+\frac{(g^{\mu\nu}(p_1-p_2)_\sigma+g_\sigma^{\nu}(p_1+2p_2)^\mu-g_\sigma^{\mu}(2p_1+p_2)^\nu)(\epsilon^{\rho\eta\sigma\chi}k_{\eta}(p_1+p_2)_\chi)}{(p_1+p_2)^2}\nn\\
&&\left.
-\epsilon^{\mu\nu\rho\sigma}(p_1+p_2-k)_\sigma \right] \,.
\eea

\noindent where $p_1,p_2$ are the initial momenta, $k$ the momenta of the gluon, and $p_3,p_4$ the momenta of the DM particle/antiparticle, $g_s$ is the SU(3) gauge coupling and
$T^a$ are the SU(3) generators in the
fundamental representation.

The corresponding squared amplitudes, averaged over initial states (color and spin) and summed over the final states are 
\bea
\sum\overline{|{\cal M}_{D1'}|^2}&=&
\frac{16}{9}\frac{ g_s^2}{\Lambda^4}
\frac{[(p_3\cdot p_4)-m_{\rm DM}^2]\left[(k\cdot(p_1+p_2))^2-2(p_1\cdot p_2)(k\cdot p_1+k\cdot p_2-p_1\cdot p_2)\right]}
{(k \cdot p_1)(k\cdot p_2)}\,\label{M2D1}\,,\nn\\ &&\\
\sum\overline{|{\cal M}_{D4'}|^2}&=&
\frac{16}{9}\frac{ g_s^2}{\Lambda^4}
\frac{[(p_3\cdot p_4)+m_{\rm DM}^2]\left[(k\cdot(p_1+p_2))^2-2(p_1\cdot p_2)(k\cdot p_1+k\cdot p_2-p_1\cdot p_2)\right]}
{(k \cdot p_1)(k\cdot p_2)}\,,\nn\\ &&
\eea
\bea
\sum\overline{|{\cal M}_{D5}|^2}&=&
-\frac{32}{9}\frac{g_s^2}{\Lambda^4}\left[\frac{(k\cdot p_1)\left[(k\cdot p_1)+(k\cdot p_2)-3(p_1\cdot p_2)-m_{\rm DM}^2 \right]}{(k\cdot p_2)}\right.\nn\\
&&\hspace{-0.5cm} +\frac{(k\cdot p_2)\left[(k\cdot p_1)+(k\cdot p_2)-3(p_1\cdot p_2)-m_{\rm DM}^2 \right]}{(k \cdot p_1)}-4(p_1\cdot p_2)\nn\\
&&\hspace{-0.5cm} { -2\frac{(p_1\cdot p_2)}{(k\cdot p_1)(k\cdot p_2)}\left[(k\cdot p_3)\left((p_1\cdot p_3)+(p_2\cdot p_3)\right)+(p_1\cdot p_2)\left(m_{\rm DM}^2+(p_1\cdot p_2)\right) \right.}\nn\\
&&\left.-2(p_1\cdot p_3)(p_2\cdot p_3)\right] \nn\\ 
&&\hspace{-0.5cm} +2\frac{(k\cdot p_3)(p_1\cdot p_3)-(p_2\cdot p_3)(p_1\cdot p_3)+(p_2\cdot p_3)^2+2(p_1\cdot p_2)^2+m_{\rm DM}^2(p_1\cdot p_2)}{(k\cdot p_2)}\nn\\
&&\hspace{-0.5cm} \left.
+2\frac{(k\cdot p_3)(p_2\cdot p_3)-(p_1\cdot p_3)(p_2\cdot p_3)+(p_1\cdot p_3)^2+2(p_1\cdot p_2)^2+m_{\rm DM}^2(p_1\cdot p_2)}{(k \cdot p_1)}\right]\,,\nn\\
\\
\sum\overline{|{\cal M}_{D8}|^2}&=&
\frac{32}{9}\frac{g_s^2}{\Lambda^4}\left[
\frac{(k\cdot p_1)\left[(k\cdot p_1)+(k\cdot p_2)-3(p_1\cdot p_2)+m_{\rm DM}^2+2(p_3\cdot p_4) \right]}{(k\cdot p_2)}\right.\nn\\
&&\hspace{-0.5cm} +\frac{(k\cdot p_2)\left[(k\cdot p_1)+(k\cdot p_2)-3(p_1\cdot p_2)+m_{\rm DM}^2+2(p_3\cdot p_4) \right]}{(k \cdot p_1)}-4(p_1\cdot p_2)\nn\\
&&\hspace{-0.5cm}  +2\frac{(p_1\cdot p_2)}{(k\cdot p_1)(k\cdot p_2)} \left[(p_1\cdot p_2)\left(2(p_3\cdot p_4)+m_{\rm DM}^2\right) +(k\cdot p_3)\left((p_1\cdot p_3)+(p_2\cdot p_3)\right) \right.\nn\\
&&\left. +2(p_1\cdot p_3)(p_2\cdot p_3) -(p_1\cdot p_2)^2 \right] \nn\\ 
&&\hspace{-0.5cm} +2\frac{(p_1\cdot p_3)\left[-(k\cdot p_3)+(p_2\cdot p_3)\right]-(p_2\cdot p_3)^2+(p_1\cdot p_2)\left[2(p_1\cdot p_2)-m_{\rm DM}^2-2(p_3\cdot p_4)\right]}{(k\cdot p_2)}\nn\\
&&\hspace{-0.5cm} \left.
+2\frac{(p_2\cdot p_3)\left[-(k\cdot p_3)+(p_1\cdot p_3)\right]-(p_1\cdot p_3)^2+(p_1\cdot p_2)\left[2(p_1\cdot p_2)-m_{\rm DM}^2-2(p_3\cdot p_4)\right]}{(k \cdot p_1)}\right]\,,\nn\\
\\
\sum\overline{|{\cal M}_{D9}|^2}&=&
\frac{128}{9}\frac{g_s^2}{\Lambda^4}
\left[-2[m_{\rm DM}^2-(k\cdot p_3)]
+\frac{(k\cdot p_1)\left[-(k\cdot p_3)+(p_1\cdot p_3)-(p_2\cdot p_3)+m_{\rm DM}^2\right]}{(k\cdot p_2)}\right.\nn \\
&&-2\frac{(p_1\cdot p_2)\left[-2(k\cdot p_3)+(p_1\cdot p_3)+(p_2\cdot p_3)+m_{\rm DM}^2\right]}{(k\cdot p_2)}\nn\\
&&-4\frac{\left[(k\cdot p_3)-(p_2\cdot p_3)\right]\left[(p_1\cdot p_3)-(p_2\cdot p_3)\right]}{(k\cdot p_2)}
\nn\\
&&+\frac{(k\cdot p_2)\left[-(k\cdot p_3)+(p_2\cdot p_3)-(p_1\cdot p_3)+m_{\rm DM}^2\right]}{(k\cdot p_1)}
\nn\\
&&-2\frac{(p_1\cdot p_2)\left[-2(k\cdot p_3)+(p_1\cdot p_3)+(p_2\cdot p_3)+m_{\rm DM}^2\right]}{(k\cdot p_1)}\nn\\
&&-4\frac{\left[(k\cdot p_3)-(p_1\cdot p_3)\right]\left[(p_2\cdot p_3)-(p_1\cdot p_3)\right]}{(k\cdot p_1)}
\nn\\
&&-2\frac{(p_1\cdot p_2)\left[(k\cdot p_3)-(p_1\cdot p_3)-(p_2\cdot p_3)\right]\left[2(k\cdot p_3)+(p_1\cdot p_2)\right]}{(k\cdot p_1)(k\cdot p_2)}\nn\\
&& \left.
+2\frac{(p_1\cdot p_2)\left[-4(p_1\cdot p_3)(p_2\cdot p_3)+m_{\rm DM}^2(p_1\cdot p_2)\right]}{(k\cdot p_1)(k\cdot p_2)}\right]\,,
\label{M2D9}
\eea
\bea
\sum\overline{|{\cal M}_{D11}|^2}&=&
\frac{3}{32\pi^2}\frac{g_s^6}{\Lambda^6}
\left[(p_3 \cdot p_4)-m_{\rm DM}^2\right]
\left\{\frac{(k\cdot p_1)^{3}}{(k\cdot p_2)(p_1\cdot p_2)}+\frac{(k\cdot p_2)^{3}}{(k\cdot p_1)(p_1\cdot p_2)}+\frac{(p_1\cdot p_2)^{3}}{(k\cdot p_1)(k\cdot p_2)}\right.\nn \\
&&+3\frac{(k\cdot p_1)(k\cdot p_2)}{(p_1\cdot p_2)}+\frac{(k\cdot p_1)(p_1\cdot p_2)-(k\cdot p_1)^2}{(k\cdot p_2)}+\frac{(k\cdot p_2)(p_1\cdot p_2)-(k\cdot p_2)^2}{(k\cdot p_1)}\nn\\
&&-\frac{(k_{-}\cdot p_1)(k\cdot p_2)^3}{(k\cdot k_{-})(k\cdot p_1)(p_1\cdot p_2)}-\frac{(k_{-}\cdot p_2)(k\cdot p_1)^3}{(k\cdot k_{-})(k\cdot p_2)(p_1\cdot p_2)}
\nn\\
&&+\frac{(k_{-}\cdot p_1)}{(k\cdot k_{-})(p_1\cdot p_2)}[(k\cdot p_1)^2+(k\cdot p_1)(k\cdot p_2)-(k\cdot p_2)^2]
\nn\\
&&+\frac{(k_{-}\cdot p_2)}{(k\cdot k_{-})(p_1\cdot p_2)}[(k\cdot p_2)^2+(k\cdot p_1)(k\cdot p_2)-(k\cdot p_1)^2]
\nn\\
&&+2\frac{(k_{-}\cdot p_1)}{(k\cdot k_{-})(k\cdot p_1)}[(k\cdot p_2)^2-(p_1\cdot p_2)(k\cdot p_2)]
\nn\\
&&+2\frac{(k_{-}\cdot p_2)}{(k\cdot k_{-})(k\cdot p_2)}[(k\cdot p_1)^2-(p_1\cdot p_2)(k\cdot p_1)]
\nn\\
&&+2\frac{(k_{-}\cdot p_1)}{(k\cdot k_{-})}[(p_1\cdot p_2)+(k\cdot p_1)-2(k\cdot p_2)]
\nn\\
&&+2\frac{(k_{-}\cdot p_2)}{(k\cdot k_{-})}[(p_1\cdot p_2)+(k\cdot p_2)-2(k\cdot p_1)]
\nn\\
&& \left.
+(k\cdot p_1)+(k\cdot p_2)+6(p_1\cdot p_2)
\right\}\,,
\label{M2D11}
\eea
\bea
\sum\overline{|{\cal M}_{D12}|^2}&=&
\frac{3}{32\pi^2}\frac{g_s^6}{\Lambda^6}
\left[(p_3 \cdot p_4)+m_{\rm DM}^2\right]
\left\{\frac{(k\cdot p_1)^{3}}{(k\cdot p_2)(p_1\cdot p_2)}+\frac{(k\cdot p_2)^{3}}{(k\cdot p_1)(p_1\cdot p_2)}+\frac{(p_1\cdot p_2)^{3}}{(k\cdot p_1)(k\cdot p_2)}\right.\nn \\
&&+3\frac{(k\cdot p_1)(k\cdot p_2)}{(p_1\cdot p_2)}+\frac{(k\cdot p_1)(p_1\cdot p_2)-(k\cdot p_1)^2}{(k\cdot p_2)}+\frac{(k\cdot p_2)(p_1\cdot p_2)-(k\cdot p_2)^2}{(k\cdot p_1)}\nn\\
&&-\frac{(k_{-}\cdot p_1)(k\cdot p_2)^3}{(k\cdot k_{-})(k\cdot p_1)(p_1\cdot p_2)}-\frac{(k_{-}\cdot p_2)(k\cdot p_1)^3}{(k\cdot k_{-})(k\cdot p_2)(p_1\cdot p_2)}
\nn\\
&&+\frac{(k_{-}\cdot p_1)}{(k\cdot k_{-})(p_1\cdot p_2)}[(k\cdot p_1)^2+(k\cdot p_1)(k\cdot p_2)-(k\cdot p_2)^2]
\nn\\
&&+\frac{(k_{-}\cdot p_2)}{(k\cdot k_{-})(p_1\cdot p_2)}[(k\cdot p_2)^2+(k\cdot p_1)(k\cdot p_2)-(k\cdot p_1)^2]
\nn\\
&&+2\frac{(k_{-}\cdot p_1)}{(k\cdot k_{-})(k\cdot p_1)}[(k\cdot p_2)^2-(p_1\cdot p_2)(k\cdot p_2)]
\nn\\
&&+2\frac{(k_{-}\cdot p_2)}{(k\cdot k_{-})(k\cdot p_2)}[(k\cdot p_1)^2-(p_1\cdot p_2)(k\cdot p_1)]
\nn\\
&&+2\frac{(k_{-}\cdot p_1)}{(k\cdot k_{-})}[(p_1\cdot p_2)+(k\cdot p_1)-2(k\cdot p_2)]
\nn\\
&&+2\frac{(k_{-}\cdot p_2)}{(k\cdot k_{-})}[(p_1\cdot p_2)+(k\cdot p_2)-2(k\cdot p_1)]
\nn\\
&& \left.
+(k\cdot p_1)+(k\cdot p_2)+6(p_1\cdot p_2)
\right\}\,,
\label{M2D12}
\eea
\bea
\sum\overline{|{\cal M}_{D13}|^2}&=&
\frac{3}{32\pi^2}\frac{g_s^6}{\Lambda^6}
\left[(p_3 \cdot p_4)-m_{\rm DM}^2\right]
\left\{\frac{(k\cdot p_1)^{3}}{(k\cdot p_2)(p_1\cdot p_2)}+\frac{(k\cdot p_2)^{3}}{(k\cdot p_1)(p_1\cdot p_2)}+\frac{(p_1\cdot p_2)^{3}}{(k\cdot p_1)(k\cdot p_2)}\right.\nn \\
&&+3\frac{(k\cdot p_1)(k\cdot p_2)}{(p_1\cdot p_2)}+\frac{(k\cdot p_1)(p_1\cdot p_2)-(k\cdot p_1)^2}{(k\cdot p_2)}+\frac{(k\cdot p_2)(p_1\cdot p_2)-(k\cdot p_2)^2}{(k\cdot p_1)}\nn\\
&&-\frac{(k_{-}\cdot p_1)(k\cdot p_2)^3}{(k\cdot k_{-})(k\cdot p_1)(p_1\cdot p_2)}-\frac{(k_{-}\cdot p_2)(k\cdot p_1)^3}{(k\cdot k_{-})(k\cdot p_2)(p_1\cdot p_2)}
\nn\\
&&+\frac{(k_{-}\cdot p_1)}{(k\cdot k_{-})(p_1\cdot p_2)}[(k\cdot p_1)^2-3(k\cdot p_1)(k\cdot p_2)+3(k\cdot p_2)^2]
\nn\\
&&+\frac{(k_{-}\cdot p_2)}{(k\cdot k_{-})(p_1\cdot p_2)}[(k\cdot p_2)^2-3(k\cdot p_1)(k\cdot p_2)+3(k\cdot p_1)^2]
\nn\\
&&+2\frac{(k_{-}\cdot p_1)}{(k\cdot k_{-})(k\cdot p_1)}[(k\cdot p_2)^2-(p_1\cdot p_2)(k\cdot p_2)]
\nn\\
&&+2\frac{(k_{-}\cdot p_2)}{(k\cdot k_{-})(k\cdot p_2)}[(k\cdot p_1)^2-(p_1\cdot p_2)(k\cdot p_1)]
\nn\\
&&+2\frac{(k_{-}\cdot p_1)}{(k\cdot k_{-})}[(p_1\cdot p_2)+(k\cdot p_1)-2(k\cdot p_2)]
\nn\\
&&+2\frac{(k_{-}\cdot p_2)}{(k\cdot k_{-})}[(p_1\cdot p_2)+(k\cdot p_2)-2(k\cdot p_1)]
\nn\\
&& \left.
-3(k\cdot p_1)-3(k\cdot p_2)+2(p_1\cdot p_2)
\right\}\,,
\label{M2D13}
\eea
\bea
\sum\overline{|{\cal M}_{D14}|^2}&=&
\frac{3}{32\pi^2}\frac{g_s^6}{\Lambda^6}
\left[(p_3 \cdot p_4)+m_{\rm DM}^2\right]
\left\{\frac{(k\cdot p_1)^{3}}{(k\cdot p_2)(p_1\cdot p_2)}+\frac{(k\cdot p_2)^{3}}{(k\cdot p_1)(p_1\cdot p_2)}+\frac{(p_1\cdot p_2)^{3}}{(k\cdot p_1)(k\cdot p_2)}\right.\nn \\
&&+3\frac{(k\cdot p_1)(k\cdot p_2)}{(p_1\cdot p_2)}+\frac{(k\cdot p_1)(p_1\cdot p_2)-(k\cdot p_1)^2}{(k\cdot p_2)}+\frac{(k\cdot p_2)(p_1\cdot p_2)-(k\cdot p_2)^2}{(k\cdot p_1)}\nn\\
&&-\frac{(k_{-}\cdot p_1)(k\cdot p_2)^3}{(k\cdot k_{-})(k\cdot p_1)(p_1\cdot p_2)}-\frac{(k_{-}\cdot p_2)(k\cdot p_1)^3}{(k\cdot k_{-})(k\cdot p_2)(p_1\cdot p_2)}
\nn\\
&&+\frac{(k_{-}\cdot p_1)}{(k\cdot k_{-})(p_1\cdot p_2)}[(k\cdot p_1)^2-3(k\cdot p_1)(k\cdot p_2)+3(k\cdot p_2)^2]
\nn\\
&&+\frac{(k_{-}\cdot p_2)}{(k\cdot k_{-})(p_1\cdot p_2)}[(k\cdot p_2)^2-3(k\cdot p_1)(k\cdot p_2)+3(k\cdot p_1)^2]
\nn\\
&&+2\frac{(k_{-}\cdot p_1)}{(k\cdot k_{-})(k\cdot p_1)}[(k\cdot p_2)^2-(p_1\cdot p_2)(k\cdot p_2)]
\nn\\
&&+2\frac{(k_{-}\cdot p_2)}{(k\cdot k_{-})(k\cdot p_2)}[(k\cdot p_1)^2-(p_1\cdot p_2)(k\cdot p_1)]
\nn\\
&&+2\frac{(k_{-}\cdot p_1)}{(k\cdot k_{-})}[(p_1\cdot p_2)+(k\cdot p_1)-2(k\cdot p_2)]
\nn\\
&&+2\frac{(k_{-}\cdot p_2)}{(k\cdot k_{-})}[(p_1\cdot p_2)+(k\cdot p_2)-2(k\cdot p_1)]
\nn\\
&& \left.
-3(k\cdot p_1)-3(k\cdot p_2)+2(p_1\cdot p_2)
\right\}\,.
\label{M2D14}
\eea
where the polarization 4-vector 
is defined as
$k_-\equiv {P(k^\nu)}/{\sqrt{k^\mu\cdot P(k_\mu)}}$, where $P$ is the parity operation.

\subsection{Cross sections}

Now, the next step is to compute the cross sections in the lab frame.
To this end we  proceed by  first evaluating the matrix elements and the phase
space density  in the center-of-mass frame
and then boosting the result to the lab frame. 
In the center-of-mass (c.o.m) frame, let us parametrize the four-momenta inolved in the process
as
\bea
p_1&=&x \frac{\sqrt{s}}{2}(1,0,0,1)\,,\qquad 
p_2=x \frac{\sqrt{s}}{2}(1,0,0,-1)\,, \qquad
k= x\frac{\sqrt{s}}{2}(z_0,z_0\hat k)\, ,\\
p_3&=&x \frac{\sqrt{s}}{2}(1-y_0,\sqrt{(1-y_0)^2-a^2}\hat p_3 )\,,\quad
p_4=x \frac{\sqrt{s}}{2}(1+y_0-z_0,\sqrt{(1+y_0-z_0)^2-a^2}\hat p_4)\, ,
\nn
\eea
where the two colliding partons  carry equal momentum fractions $x_1=x_2\equiv x$ of the incoming
protons, $a\equiv 2 m_{\rm DM}/(x\sqrt{s})<1$, $\hat k=(0,\sin\theta_0,\cos\theta_0)$, and 
$\theta_0$ is the polar angle of $\hat k$ with respect to the beam line, in the c.o.m. frame.
With the subscript $_0$ we will  refer to quantities evaluated in the c.o.m. frame.
The polarization 4-vector $k_-$ in the c.o.m. frame simply reads $k_-=(1/\sqrt{2})(1,0,-\sin\theta_0,-\cos\theta_0)$.

The  conservation of three-momentum sets the angle $\theta_{0\,3j}$
 between $\hat p_3$ and $\hat k$ as:
$ \cos\theta_{0\,3j}=(\mathbf{p}_4^2-\mathbf{k}^2-\mathbf{p}_3^2)/{2|\mathbf{k}||\mathbf{p}_3|}$.
For the doubly-differential cross sections with respect to the energy and angle  of the emitted gluon, 
in the c.o.m. frame, we obtain
\bea
\left.\frac{\de^2\hat\sigma}{\de z_0\de\cos\theta_0}\right\vert_{D1'}
&=&
\frac{\alpha_s}{36\pi^2}\frac{x^2 s}{\Lambda^4}
\frac{\left[1-z_0-\frac{4 m_{\rm DM}^2}{x^2 s}\right]^{3/2}}{\sqrt{1-z_0}}\frac{[1+(1-z_0)^2]}{z_0\sin^2\theta_0}\,,  
\label{d2sigmaD1}\\
\left.\frac{\de^2\hat\sigma}{\de z_0\de\cos\theta_0}\right\vert_{D4'}
&=&
\frac{\alpha_s}{36\pi^2}\frac{x^2 s}{\Lambda^4}
\frac{\left[1-z_0-\frac{4 m_{\rm DM}^2}{x^2 s}\right]^{1/2}}{\sqrt{1-z_0}}\frac{[1+(1-z_0)^2]}{z_0\sin^2\theta_0}\,, \\
\left.\frac{\de^2\hat\sigma}{\de z_0\de\cos\theta_0}\right\vert_{D5} 
&=&
\frac{\alpha_s}{108\pi^2}\frac{x^2 s}{\Lambda^4}
\frac{\sqrt{1-z_0-\frac{4 m_{\rm DM}^2}{x^2 s}}}{\sqrt{1-z_0}}\frac{(1-z_0+\frac{2 m_{\rm DM}^2}{x^2 s})(8-8z_0+(3+\cos2\theta_0)z_0^2)}{z_0\sin^2\theta_0} \,, \\
\left.\frac{\de^2\hat\sigma}{\de z_0\de\cos\theta_0}\right\vert_{D8}
&=&
\frac{\alpha_s}{108\pi^2}\frac{x^2 s}{\Lambda^4}
\frac{[1-z_0-\frac{4 m_{\rm DM}^2}{x^2 s}]^{3/2}}{\sqrt{1-z_0}}\frac{8-8z_0+(3+\cos2\theta_0)z_0^2}{z_0\sin^2\theta_0} \,, \label{d2sigmaD8}\\
\left.\frac{\de^2\hat\sigma}{\de z_0\de\cos\theta_0}\right\vert_{D9}
&=&
\frac{\alpha_s}{27\pi^2}\frac{x^2 s}{\Lambda^4}
\frac{\sqrt{1-z_0-\frac{4 m_{\rm DM}^2}{x^2 s}}}{[1-z_0]^{3/2}}\frac{(1-z_0+\frac{2 m_{\rm DM}^2}{x^2 s})(4-8z_0+6z_0^2-(1+\cos2\theta_0)z_0^3)}{z_0\sin^2\theta_0}\,, \label{d2sigmaD9}\nn\\
&&\\
\left.\frac{\de^2\hat\sigma}{\de z_0\de\cos\theta_0}\right\vert_{D11}
&=&
\frac{3\alpha_s^3 x^4 s^2}{32768\pi^2\Lambda^6}
\frac{\left[1-z_0-\frac{4 m_{\rm DM}^2}{x^2 s}\right]^{3/2}}{z_0\sqrt{1-z_0}\sin^2\theta_0}\left[128-128(1+\cos2\theta_0)z_0\right.\nn\\
&&
+ (304+64\cos2\theta_0+16\cos4\theta_0)z_0^2-128(1+\cos2\theta_0)z_0^3\nn\\
&&\left. +(79+44\cos2\theta_0+5\cos4\theta_0)z_0^4\right] \,, \label{d2sigmaD11} \\
\left.\frac{\de^2\hat\sigma}{\de z_0\de\cos\theta_0}\right\vert_{D12}
&=&
\frac{3\alpha_s^3 x^4 s^2}{32768\pi^2\Lambda^6}
\frac{\sqrt{1-z_0-\frac{4 m_{\rm DM}^2}{x^2 s}}\sqrt{1-z_0}}{z_0\sin^2\theta_0}\left[128-128(1+\cos2\theta_0)z_0\right.\nn\\
&&
+ (304+64\cos2\theta_0+16\cos4\theta_0)z_0^2-128(1+\cos2\theta_0)z_0^3\nn\\
&& \left.+(79+44\cos2\theta_0+5\cos4\theta_0)z_0^4\right] \,, \label{d2sigmaD12} 
\eea
\bea
\left.\frac{\de^2\hat\sigma}{\de z_0\de\cos\theta_0}\right\vert_{D13}
&=&
\frac{3\alpha_s^3 x^4 s^2}{32768\pi^2\Lambda^6}
\frac{\left[1-z_0-\frac{4 m_{\rm DM}^2}{x^2 s}\right]^{3/2}}{z_0\sqrt{1-z_0}\sin^2\theta_0}\left[128-128(1+\cos2\theta_0)z_0\right.\nn\\
&&
+ (240+128\cos2\theta_0+16\cos4\theta_0)z_0^2-16(11+4\cos2\theta_0+\cos4\theta_0)z_0^3\nn\\
&&\left.+(79+44\cos2\theta_0+5\cos4\theta_0)z_0^4\right] \,, \label{d2sigmaD13} \\
\left.\frac{\de^2\hat\sigma}{\de z_0\de\cos\theta_0}\right\vert_{D14}
&=&
\frac{3\alpha_s^3 x^4 s^2}{32768\pi^2\Lambda^6}
\frac{\sqrt{1-z_0-\frac{4 m_{\rm DM}^2}{x^2 s}}\sqrt{1-z_0}}{z_0\sin^2\theta_0}\left[128-128(1+\cos2\theta_0)z_0\right.\nn\\
&&
+ (240+128\cos2\theta_0+16\cos4\theta_0)z_0^2-16(11+4\cos2\theta_0+\cos4\theta_0)z_0^3\nn\\
&&\left.+(79+44\cos2\theta_0+5\cos4\theta_0)z_0^4\right] \,. \label{d2sigmaD14}
\eea
Eq.~(\ref{d2sigmaD1})-(\ref{d2sigmaD8})  agree with the findings  in Refs.~\cite{C, D}, up to the factor of 1/9, as we are considering colored colliding particles.

To get the cross sections in the lab frame we perform a 
 boost along the $\hat z$-axis, accounting for generic parton momentum fractions $x_1, x_2$.
Also, the energy and angle of the emitted gluon are translated into momentum transfer $p_{\rm T}$
and pseudo-rapidity $\eta$.
This way we get the translation of Eqs.~(\ref{d2sigmaD1})-(\ref{d2sigmaD9}) into  the lab frame
\bea
\left.\frac{\de^2\hat\sigma}{\de p_{\rm T}\de\eta}\right\vert_{D1'}
&=&\frac{ \alpha_s}{36 \pi^2}\frac{x_1x_2 s}{\Lambda^4}
\frac{1}{p_{\rm T}}
\frac{\left[1-f-\frac{4m_{\rm DM}^2}{x_1x_2 s}\right]^{3/2}\left[1+\left(1-f\right)^2\right]}{\sqrt{1-f}}\,,
\label{d2sigmalabD1} \\
\left.\frac{\de^2\hat\sigma}{\de p_{\rm T}\de\eta}\right\vert_{D4'}
&=&\frac{ \alpha_s}{36 \pi^2}\frac{x_1x_2 s}{\Lambda^4}
\frac{\sqrt{1-f}}{p_{\rm T}}
\left[1-f-\frac{4m_{\rm DM}^2}{x_1x_2 s}\right]^{1/2}\left[1+\left(1-f\right)^2\right]\,, \\
\left.\frac{\de^2\hat\sigma}{\de p_{\rm T}\de\eta}\right\vert_{D5}
&=&
\frac{\alpha_s}{27\pi^2}\frac{x_1x_2 s}{\Lambda^4}
\frac{\sqrt{1-f-\frac{4 m_{\rm DM}^2}{x_1x_2 s}}}{\sqrt{1-f}}\frac{\left[1-f+\frac{2 m_{\rm DM}^2}{x_1x_2 s}\right]
\left[1+\left(1-f\right)^2-2\frac{p_{\rm T}^2}{x_1x_2 s}\right]}{p_{\rm T}}\,, \\
\left.\frac{\de^2\hat\sigma}{\de p_{\rm T}\de\eta}\right\vert_{D8}
&=&
\frac{\alpha_s}{27\pi^2}\frac{x_1x_2 s}{\Lambda^4}
\frac{[1-f-\frac{4 m_{\rm DM}^2}{x_1x_2 s}]^{3/2}}{\sqrt{1-f}}\frac{1+\left(1-f\right)^2-2\frac{p_{\rm T}^2}{x_1x_2 s}}{p_{\rm T}}\,,\\
\left.\frac{\de^2\hat\sigma}{\de p_{\rm T}\de\eta}\right\vert_{D9}
&=&
\frac{2\alpha_s}{27\pi^2}\frac{x_1 x_2 s}{\Lambda^4}
\frac{\sqrt{1-f-\frac{4 m_{\rm DM}^2}{s x_1 x_2}}}{[1-f]^{3/2}}\frac{(1-f+\frac{2 m_{\rm DM}^2}{x_1 x_2 s})\left[(1-f)(1+(1-f)^2)+f\frac{4p_{\rm T}^2}{ x_1 x_2 s}\right]}{p_{\rm T}}\,, \nn\\
&&\label{d2sigmalabD9}\\
\left.\frac{\de^2\hat\sigma}{\de p_{\rm T}\de\eta}\right\vert_{D11}
&=&
\frac{3\alpha_s^3 x_1^2 x_2^2 s^2}{256\pi^2\Lambda^6}
\frac{(1-f-\frac{4 m_{\rm DM}^2}{s x_1 x_2})^{3/2}}{p_{\rm T}f^2\sqrt{1-f}}\left[16\frac{p_{\rm T}^4}{ x_1^2 x_2^2 s^2}+8\frac{p_{\rm T}^2}{ x_1 x_2 s}f+(1-8\frac{p_{\rm T}^2}{ x_1 x_2 s}+5\frac{p_{\rm T}^4}{ x_1^2 x_2^2 s^2})f^2\right.\nn\\
&&
+\left. (-2+8\frac{p_{\rm T}^2}{ x_1 x_2 s})f^3+(3-4\frac{p_{\rm T}^2}{ x_1 x_2 s})f^4-2f^5+f^6\right]\,, \label{d2sigmalabD11}\\
\left.\frac{\de^2\hat\sigma}{\de p_{\rm T}\de\eta}\right\vert_{D12}
&=&
\frac{3\alpha_s^3 x_1^2 x_2^2 s^2}{256\pi^2\Lambda^6}
\frac{\sqrt{1-f-\frac{4 m_{\rm DM}^2}{s x_1 x_2}}\sqrt{1-f}}{p_{\rm T}f^2}\left[16\frac{p_{\rm T}^4}{ x_1^2 x_2^2 s^2}+8\frac{p_{\rm T}^2}{ x_1 x_2 s}f\right.\nn\\
&&
\left. 
+(1-8\frac{p_{\rm T}^2}{ x_1 x_2 s}+5\frac{p_{\rm T}^4}{ x_1^2 x_2^2 s^2})f^2
+(-2+8\frac{p_{\rm T}^2}{ x_1 x_2 s})f^3+(3-4\frac{p_{\rm T}^2}{ x_1 x_2 s})f^4-2f^5+f^6\right]\,, \nn\\
&&\label{d2sigmalabD12}
\eea
\bea
\left.\frac{\de^2\hat\sigma}{\de p_{\rm T}\de\eta}\right\vert_{D13}
&=&
\frac{3\alpha_s^3 x_1^2 x_2^2 s^2}{256\pi^2\Lambda^6}
\frac{(1-f-\frac{4 m_{\rm DM}^2}{s x_1 x_2})^{3/2}}{p_{\rm T}f^2\sqrt{1-f}}\left[16\frac{p_{\rm T}^4}{ x_1^2 x_2^2 s^2}+8(\frac{p_{\rm T}^2}{ x_1 x_2 s}-2\frac{p_{\rm T}^4}{ x_1^2 x_2^2 s^2})f\right.
\nn\\
&&
+\left. (1-12\frac{p_{\rm T}^2}{ x_1 x_2 s}+5\frac{p_{\rm T}^4}{ x_1^2 x_2^2 s^2})f^2+(-2+8\frac{p_{\rm T}^2}{ x_1 x_2 s})f^3+(3-4\frac{p_{\rm T}^2}{ x_1 x_2 s})f^4-2f^5+f^6\right]\,, \nn\\
&&\label{d2sigmalabD13}\\
\left.\frac{\de^2\hat\sigma}{\de p_{\rm T}\de\eta}\right\vert_{D14}
&=&
\frac{3\alpha_s^3 x_1^2 x_2^2 s^2}{256\pi^2\Lambda^6}
\frac{\sqrt{1-f-\frac{4 m_{\rm DM}^2}{s x_1 x_2}}\sqrt{1-f}}{p_{\rm T}f^2}\left[16\frac{p_{\rm T}^4}{ x_1^2 x_2^2 s^2}+8(\frac{p_{\rm T}^2}{ x_1 x_2 s}-2\frac{p_{\rm T}^4}{ x_1^2 x_2^2 s^2})f\right.
\nn\\
&&
+\left. (1-12\frac{p_{\rm T}^2}{ x_1 x_2 s}+5\frac{p_{\rm T}^4}{ x_1^2 x_2^2 s^2})f^2+(-2+8\frac{p_{\rm T}^2}{ x_1 x_2 s})f^3+(3-4\frac{p_{\rm T}^2}{ x_1 x_2 s})f^4-2f^5+f^6\right]\,, \nn\\
\label{d2sigmalabD14}
\eea
where we have defined
\be
f(p_{\rm T}, \eta_,x_1,x_2)\equiv
\frac{p_{\rm T}(x_1 e^{-\eta}+x_2 e^\eta)}{x_1 x_2\sqrt{ s}}\,.
\ee
For the emission of a photon, rather than a gluon, from a quark with charge $Q_q$
one simply replaces 
$(4/3)\, \alpha_s\to Q_q^2\alpha $ in Eqs.~(\ref{d2sigmalabD1})-(\ref{d2sigmalabD9}). 
From these expressions one reproduces the results reported in Eqs.~(\ref{d2sigmaefflab1})-(\ref{d2sigmaefflab14}).

%%%%%%%%%%%%%%%%%%%%%%%%%%%%%%%%%%%%%%%%%%%%
\end{document}